\documentclass[aps,pra]{revtex4-2}
\usepackage{graphicx}
\begin{document}
\title{Superradiant atomic recoil lasing with orbital angular momentum light}
\author{A.T. Gisbert}
\affiliation{Dipartimento di Fisica "Aldo Pontremoli", Universit\`{a} degli Studi di Milano, Via Celoria 16, I-20133 Milano, Italy}
\author{N. Piovella}
\affiliation{Dipartimento di Fisica "Aldo Pontremoli", Universit\`{a} degli Studi di Milano, Via Celoria 16, I-20133 Milano, Italy}
\author{G.R.M. Robb}
\affiliation{SUPA and Department of Physics, University of Strathclyde,Glasgow G4 0NG, Scotland, UK}
\author{D.G. McLellan}
\affiliation{SUPA and Department of Physics, University of Strathclyde,Glasgow G4 0NG, Scotland, UK}

\date{\today}
\begin{abstract}
We analyze the collective light scattering by cold atoms in free space of a pump laser beam possessing orbital angular momentum. We derive a set of coupled equations for the atomic motion in which the vacuum-mode field is adiabatically eliminated. The resulting equations describe collective recoil as due to either transfer of linear momentum or orbital angular momentum. For a transverse annular atomic distribution the initial equilibrium with uniform atomic phases and no scattered field, is unstable. The atoms are set in rotation and bunched in phase at different harmonics depending on the pump azimuthal index $\ell$ and on the ring radius.
\end{abstract}

\maketitle

\section{INTRODUCTION}

Opto-mechanical collective effects in cold atomic systems, where atomic density structures and optical fields evolve simultaneously under their mutual influence, have been a topic of interest for several years now.  In particular, the exchange of (linear) momentum between the photons of an incident optical pump and those of a back-scattered field in an optical ring cavity has led to the concept of the collective atomic recoil laser (CARL), predicted in the 1990's \cite{Bonifacio1994,Bonifacio1994b} and subsequently experimentally observed \cite{kruse2003observation,Slama2007}. It is well known that light may carry also spin and orbital angular momentum (SAM \cite{Beth1935} and OAM \cite{Allen1992} respectively) in addition to linear momentum. Transfer of OAM from light to atoms was first studied theoretically in \cite{babiker1994light}, and has been studied extensively ever since \cite{Franke-Arnold2017}. The idea of an analogous effect to CARL involving collective exchange of OAM between a pump and a scattered field was proposed 
in \cite{robb2012superradiant,robb2012collective}. In this scheme, two co-propagating, counter-rotating Laguerre-Gaussian beams incident on a gas of ultracold atoms give rise to a superradiant instability in which orbital angular momentum is transferred to the atoms and the probe field is amplified. The atoms acquire angular momentum in discrete amounts of $2\ell\hbar$, where $\ell$ is the magnitude of the azimuthal index of the Laguerre-Gaussian beams. The instability involves the development of an azimuthal density modulation in the atomic distribution. Since the pump and probe beams are copropagating, no net exchange of linear momentum occurs in the scattering process. Similar theoretical studies of superradiant scattering into radially propagating end-fire modes from a pancake shaped Bose-Einstein Condensate (BEC) were performed in \cite{Tasgin_2011,Das_2016}, where the production of vortices in the BEC was predicted as a consequence of the superradiant scattering process.

Recently, a multi-mode theory of CARL in free space has been developed \cite{Ayllon2019,gisbert2020multimode}, where no assumption about the scattered direction was made. Contrary to the usual CARL phenomena, where the light scattering occurs effectively in a single spatial mode defined by the axis of a ring cavity or the major axis of an elongated atomic sample, in free space the pump photons are initially scattered into the 3D vacuum modes, and only at later times the dominant modes of the scattered light emerge naturally from the collective process. 

In this paper, we extend the multi-mode CARL theory for a pump laser consisting of a hollow Laguerre-Gaussian beam with azimuthal index $\ell$, without any assumption about the scattered field. The model leads to a set of coupled equations for the atoms in which the scattered field is adiabatically eliminated. This model describes a superradiant emission process involving exchange of either linear momentum, in units of $\mathbf{p}=\hbar \mathbf{k}$, or orbital angular momentum (OAM), in units of $\hbar\ell$. 
In particular, we investigate the dynamics of the atomic motion for an annular transverse distribution in which the longitudinal motion along the pump  direction is neglected. We find that the initial equilibrium, with atoms uniformly distributed along the ring and no photons, is unstable. The atoms are set in rotation and bunched in phase by a collective process, emitting a field with a transverse pattern distribution depending on the azimuthal index $\ell$ and on the ring radius.

The paper is organized as follow. In sec. II we present the theoretical model, starting from the Hamiltonian of $N$ two-level atoms interacting with the OAM pump and the vacuum-mode field. We derive from it the single-atom forces, the OAM-CARL multimode equations, the superradiant equations describing the 3D atomic motion, and the scattered field. In sec. III we consider the motion in a transverse plane neglecting the longitudinal motion. Sec.IV further reduces the atomic motion to the case where the atoms are trapped in a thin circular ring. In Sec.V we address the one-dimensional dynamics of the atoms in a ring, calculating the equilibrium and the linear stability. Numerical results for the nonlinear evolution are presented in Sec. VI and conclusions are drawn in Sec. VII.

\section{MODEL}

\begin{figure}[hbt!]
\centerline{\includegraphics[width=0.6\textwidth]{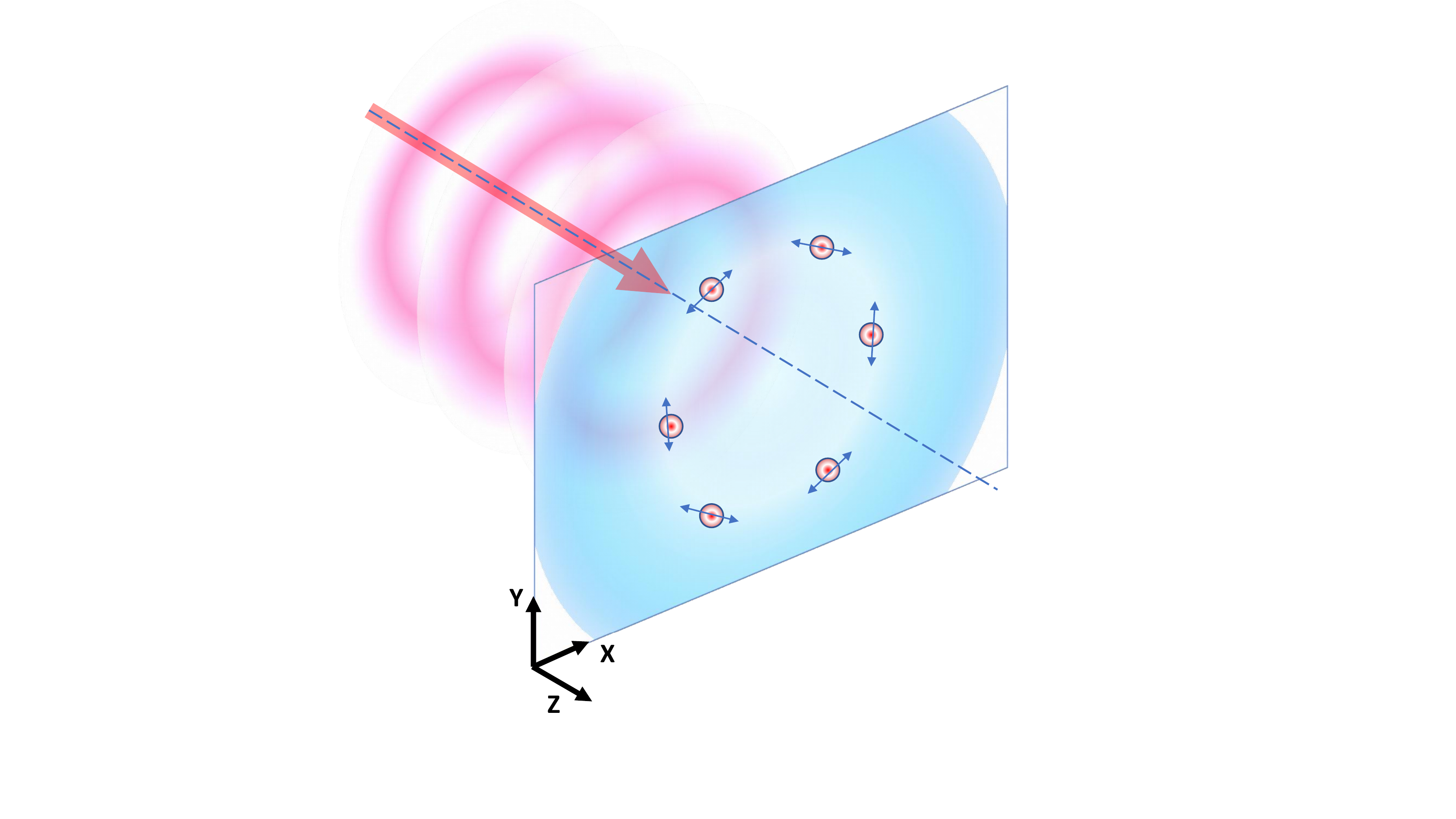}}
\caption{Schematic of the experimental setup. The pump beam is incident along the $z$ axis on an annular atomic distribution.}
\label{setup}
\end{figure}

A schematic representation of the setup is shown in fig.\ref{setup}. It consists of a collection of cold atoms in an annular configuration \cite{Amico2005,Franke-Arnold2007} in the plane $(x,y)$, upon which a pump field consisting of a Laguerre-Gaussian mode is incident along the $z$ direction. The atoms scatter the incident photon in all the 3D direction. In the following we introduce the equations describing the evolution of the center-of-mass of each atom under the optical dipole force of the pump and scattered field, and of the multimode scattered field. In free space the optical field can be adiabatically eliminated, leading to a set of coupled equations for the atoms.

\subsection{OAM pump field}

We assume a pump consisting of a hollow Laguerre-Gaussian mode $LG_{0\ell}$ directed along the $z$-axis and linearly polarized, whose electric field component, expressed in cylindrical coordinates, has the form \cite{yao2011orbital}:
\begin{equation}
E_0(\rho,\phi,z)=A_0 R(\rho)e^{i(k_0z+\ell\phi-\omega_0t)}+\mathrm{c.c.}\label{OAM:pump}
\end{equation}
We assume, without loss of generality, $\ell>0$, and define 
\begin{eqnarray}
R(\rho)&=&e^{-\rho^2/w^2}\left(\frac{\sqrt{2}\rho}{w}\right)^{\ell},\label{R(r)}\\
A_0 &=& \sqrt{\frac{2^{\ell+1}P_0}{\pi w^2\ell!\epsilon_0 c}},
\end{eqnarray}
where $w$ is the beam waist (assumed constant) and $P_0$ is the mode power. 

\subsection{Hamiltonian}

We assume $N$ two-level atoms with internal states $|g_j\rangle$ and $|e_j\rangle$, with positions and momenta $\mathbf{r}_j$ and $\mathbf{p}_j=M\mathbf{v}_j$ (with $j=1,\dots,N$), mass $M$, resonant frequency $\omega_a$, dipole $d$, interacting with a pump OAM field given  by Eq.(\ref{OAM:pump}) and scattering photons in the vacuum modes with wavenumber $\mathbf{k}$ and frequency $\omega_k$. The system is described by the Hamiltonian
\begin{equation}
H=H_L+H_V.\label{H}
\end{equation}
The first term of Eq.(\ref{H}) describes the interaction of the atoms with the pump laser, with
\begin{eqnarray}\label{HL}
    H_L =\sum_{j=1}^N\frac{\mathbf{p}_j^2}{2M}+\frac{\hbar\Omega_0}{2}\sum_{j=1}^N R(\rho_j)\left[\sigma_{j}^- 
    e^{i\Delta_0t-ik_0z_j-i\ell\phi_j}+\sigma_{j}^+ 
    e^{-i\Delta_0t+ik_0z_j+i\ell\phi_j}\right],
\end{eqnarray}
where $\Omega_0=dA_0/\hbar$ is the pump Rabi frequency and $\Delta_0=\omega_0-\omega_a$ is the pump-atom detuning.
The internal dynamics of the two-level atoms are described by the operators
$\sigma_{j}^z=|e_j\rangle \langle e_j|- |g_j\rangle \langle g_j|$,~
$\sigma_{j}^+=|e_j\rangle\langle g_j|$ and
$\sigma_{j}^-=|g_j\rangle\langle e_j|$. The second term of Eq.(\ref{H}) describes the interaction between the atoms and the vacuum-mode field, with
\begin{eqnarray}\label{HV}
    H_V =\hbar\sum_{\mathbf{k}}g_k\left[a_{\mathbf{k}}^\dagger \sigma_{j}^- e^{i\Delta_kt-i\mathbf{k}\cdot\mathbf{r}_j}
        +\sigma_{j}^+a_{\mathbf{k}}e^{-i\Delta_kt+i\mathbf{k}\cdot\mathbf{r}_j}\right].
\end{eqnarray}
The vacuum modes, described by the operators $a_\mathbf{k}$, have wave-vectors $\mathbf{k}$ and frequency $\omega_k$ with
$\Delta_k=\omega_k-\omega_a$, coupling rate $g_k=d[\omega_k/(2\hbar\epsilon_0 V_{ph})]^{1/2}$, being $V_{ph}$ the quantization volume. We disregard polarization and short-range effects, using a scalar model for the radiation field.

The Heisenberg equations for the dipole operators $\sigma_j^-$ are:
\begin{eqnarray}
  \dot\sigma_{j}^- &=& i\left[\frac{\Omega_0}{2}
  R(\rho_j)e^{-i\Delta_0 t+ik_0z_j+i\ell\phi_j} +\sum_{\mathbf{k}}g_k
  a_{\mathbf{k}}e^{-i\Delta_k t+i \mathbf{k}\cdot \mathbf{r}_j}\right]\sigma_{j}^z.
\end{eqnarray}
Introducing $\sigma_j=\sigma_{j}^- e^{i\Delta_0 t}$ and neglecting the population of the excited state (assuming weak field and/or large detuning $\Delta_0$), so that $\sigma_{j}^z\approx -1$:
\begin{eqnarray}
  \dot\sigma_{j} &=&\left(i\Delta_0-\frac{\Gamma}{2}\right)\sigma_j- \frac{i\Omega_0}{2}R(\rho_j)e^{ik_0z_j+i\ell\phi_j}
   -i\sum_{\mathbf{k}}g_k
  a_{\mathbf{k}}e^{-i(\omega_k-\omega_0)t+i \mathbf{k}\cdot \mathbf{r}_j}\label{s1},
\end{eqnarray}
where we added the spontaneous emission decay term $-(\Gamma/2)\sigma_j$, with $\Gamma=d^2k^3/2\pi\epsilon_0\hbar$ as the spontaneous decay rate. Assuming $\Gamma\gg\omega_{\mathrm{rec}}$, being $\omega_{\mathrm{rec}}=\hbar k^2/2M$ the recoil frequency, we can  adiabatically eliminate the internal degree of freedom, taking $\dot\sigma_j\approx 0$ in Eq.(\ref{s1}):
\begin{eqnarray}
\sigma_j &\approx & \frac{1}{\Delta_0+i\Gamma/2}\left[\frac{\Omega_0}{2}R(\rho_j)e^{ik_0z_j+i\ell\phi_j}
  +\sum_{\mathbf{k}}g_k
 a_{\mathbf{k}}e^{-i(\omega_k-\omega_0)t+i\mathbf{k}\cdot \mathbf{r}_j}\right]\label{s1:adia}.
 \end{eqnarray}
The first term describes the dipole excitation induced by the driving field, whereas the second term is the excitation induced by the scattered field.

\subsection{Single-atom forces}

The OAM laser beam induces on each atom a mechanical force
\begin{equation}
\mathbf{F}_{Lj}=-\mathbf{\nabla}_j H_L=\mathbf{F}_{z j}+\mathbf{F}_{\phi j}+\mathbf{F}_{\rho j}, \label{FL}
\end{equation}
whose components can be derived using Eq.(\ref{HL}), assuming cylindrical coordinates and retaining only the first term of Eq.(\ref{s1:adia}),
\begin{eqnarray}
\mathbf{F}_{z j}&=&-\mathbf{\hat u}_z\frac{\partial H_L}{\partial z_j}=\hbar k_0\frac{\Gamma}{\Delta_0^2+\Gamma^2/4}
\left[\frac{\Omega_0R(\rho_j)}{2}\right]^2 \mathbf{\hat u}_z\label{Fz},\\
\mathbf{F}_{\phi j}&=&-\frac{\mathbf{\hat u}_\phi}{\rho_j}\frac{\partial H_L}{\partial\phi_j}=\ell\hbar\frac{\Gamma}{\Delta_0^2+\Gamma^2/4}\left[\frac{\Omega_0R(\rho_j)}{2}\right]^2 \frac{\mathbf{\hat u}_\phi}{\rho_j}\label{Fphi},\\
\mathbf{F}_{\rho j}&=&-\mathbf{\hat u}_\rho\frac{\partial H_L}{\partial\rho_j}=-
\frac{2\hbar\Delta_0}{\Delta_0^2+\Gamma^2/4}\left[\frac{\Omega_0R(\rho_j)}{2}\right]^2\left[\frac{\ell}{\rho_j}-\frac{2\rho_j}{w^2}\right]\mathbf{\hat u}_\rho\label{Frho}.
\end{eqnarray}
The force $\mathbf{F}_{z j}$ is the usual radiation pressure force, directed along the incidence direction of the laser beam, with $\mathbf{k}_0=k_0\mathbf{\hat u}_z$. The azimuthal force $\mathbf{F}_{\phi j}$  produces a torque directed along the $z$-axis, which put the atoms in rotation \cite{babiker1994light}. The radial force $\mathbf{F}_{\rho j}$ is zero for $\rho_j=w\sqrt{\ell/2}$ (where $R(\rho_j)$ is maximum) and is focusing toward this maximum for $\Delta_0<0$. 

\subsection{Multi-mode OAM-CARL model}

Assuming the pump field detuning $\Delta_0\gg \Gamma$, we neglect the multiple scattering processes.
The resulting multimode equations describing the collective recoil in the presence of an OAM pump beam are (see Appendix \ref{App:1}):
\begin{eqnarray}
\dot \mathbf{r}_j, &=& \frac{\mathbf{p}_j}{M},\label{D:r:j}\\
 \dot \mathbf{p}_{j}
 &=&i\hbar g R(\rho_j)
 \sum_{\mathbf{k}}(\mathbf{k}_0-\mathbf{k})\left\{a_{\mathbf{k}}
 e^{-i(\mathbf{k}_0-\mathbf{k})\cdot \mathbf{r}_j-i\ell\phi_j-i\delta_k t}-\textrm{h.c.}\right\}\nonumber\\
 &-&
 \hbar g
 \left\{\left[\frac{dR(\rho_j)}{d\rho_j}\mathbf{\hat u}_\rho-i\ell\frac{R(\rho_j)}{\rho_j}\mathbf{\hat u}_\phi\right]\sum_{\mathbf{k}}a_{\mathbf{k}}
 e^{-i(\mathbf{k}_0-\mathbf{k})\cdot \mathbf{r}_j-i\delta_k t}+\textrm{h.c.}\right\}+\mathbf{F}_{Lj},
 \label{force:z:j}\\
  \dot a_{\mathbf{k}} &=& -ig\sum_{j=1}^N
R(\rho_j)e^{i(\mathbf{k}_0-\mathbf{k})\cdot \mathbf{r}_j+i\ell\phi_j+i\delta_k t} \label{D:A:j},
 \end{eqnarray}
where $j=1,\dots,N$, $\delta_k=\omega_k-\omega_0$, $g=g_k(\Omega_0/2\Delta_0)$ and the sum in $\mathbf{k}$ is over all the vacuum three-dimensional modes. In the r.h.s. of Eq.(\ref{force:z:j}), the first term  is the force proportional to the photon momentum exchange $\hbar(\mathbf{k}_0-\mathbf{k})$ \cite{Ayllon2019}, whilst the second term is the collective force proportional to the gradient of the pump amplitude profile and of the azimuthal phase. In the limit $\Delta_0\gg \Gamma$ the single-atom force (\ref{FL}) takes the form:
\begin{equation}
\mathbf{F}_{Lj}=\Gamma
\left[\frac{\Omega_0R(\rho_j)}{2\Delta_0}\right]^2 
\left(\hbar k_0\mathbf{\hat u}_z+\frac{\hbar \ell}{\rho_j}\mathbf{\hat u}_\phi\right)
+
\frac{\hbar[\Omega_0R(\rho_j)]^2}{2\Delta_0}\left[\frac{\ell}{\rho_j}-\frac{2\rho_j}{w^2}\right]\mathbf{\hat u}_\rho.\label{FL:j}
\end{equation}
Eqs.(\ref{D:r:j})-(\ref{D:A:j}) extend the single-mode OAM-CARL described in ref.\cite{robb2012superradiant}, where the atoms are assumed to scatter the pump photons in a co-propagating, with $\mathbf{k}=\mathbf{k}_0$, and counter-rotating Laguerre-Gaussian mode $LG_{0-\ell}$ with the same beam waist $w$. Furthermore, ref.\cite{robb2012superradiant} assumes the atoms trapped in a thin circular ring in the transverse plane such that the radial motion is negligible, i.e., $\rho_j$ is constant and equal to $\rho=w\sqrt{\ell/2}$ (for which $R(\rho)$ is maximum). In this case the evolution of the atomic cloud is essentially one-dimensional, in the azimuthal coordinate alone.

\subsection{Superradiant OAM-CARL model}

In free space we can assume the Markov approximation and adiabatically eliminate the multimode radiation field $a_{\mathbf{k}}$ in Eqs.(\ref{D:r:j})-(\ref{D:A:j}) (see Appendix \ref{App:2}). The resulting equations are:
\begin{eqnarray}
\dot \mathbf{r}_j&=& \frac{\mathbf{p}_j}{M},\label{CARL:rj}\\
\dot \mathbf{p}_j &=&
\Gamma\hbar k_0\left(\frac{\Omega_0}{2\Delta_0}\right)^2R(\rho_j)\sum_{m\neq j}R(\rho_m)\left\{(\hat\mathbf{z}-\hat\mathbf{r}_{jm})
\frac{\sin[k_0(r_{jm}-z_{jm})-\ell\phi_{jm}]}{k_0r_{jm}}-\hat\mathbf{r}_{jm}\frac{\cos[k_0(r_{jm}-z_{jm})-\ell\phi_{jm}]}{(k_0r_{jm})^2}
\right\}\nonumber\\
&+&
\hbar\Gamma\left(\frac{\Omega_0}{2\Delta_0}\right)^2R(\rho_j)\left[\frac{\ell}{\rho_j}-\frac{2\rho_j}{w^2}\right]
\sum_{m\neq j}R(\rho_m)\frac{\cos[k_0(r_{jm}-z_{jm})-\ell\phi_{jm}]}{k_0r_{jm}}\mathbf{\hat u}_\rho\nonumber\\
&+&\ell\hbar\Gamma\left(\frac{\Omega_0}{2\Delta_0}\right)^2\frac{R(\rho_j)}{\rho_j}\sum_{m\neq j}R(\rho_m)\frac{\sin[k_0(r_{jm}-z_{jm})-\ell\phi_{jm}]}{k_0r_{jm}}\mathbf{\hat u}_\phi+\mathbf{F}_{Lj},
\label{CARL:pj}
\end{eqnarray}
where $\phi_{jm}=\phi_j-\phi_m$, $r_{jm}=|\mathbf{r}_j-\mathbf{r}_m|$ and  $\hat\mathbf{r}_{jm}=(\mathbf{r}_j-\mathbf{r}_m)/r_{jm}$, which is the unit vector along the distance between $\mathbf{r}_j$ and $\mathbf{r}_m$. The first term of the r.h.s. of Eq.(\ref{CARL:pj}) is due to the collective momentum recoil, weighted by the radial profile $R(\rho)$ of the pump field; the second term is a collective radial force, due to variation of the radial profile of the pump, and the third one is the collective azimuth force, proportional to the gradient of the OAM pump phase.
It is possible to return to the results obtained for CARL in free space in \cite{Ayllon2019}, where the atoms are driven by a plane wave, by considering the first term only, with $R(\rho_j)=1$ and $\ell=0$.

\subsection{Scattered field}

The scattered intensity in the far-field limit, for $\mathbf{r}\gg\mathbf{r}_j$, is
\begin{eqnarray}
 I_s(\mathbf{k})&=&I_1N^2 |M(\mathbf{k},t)|^2,
 \end{eqnarray}
where $I_1=(\hbar\omega_0\Gamma/8\pi r^2)(\Omega_0/2\Delta_0)^2$ is the single-atom Rayleigh scattering intensity and
\begin{eqnarray}
M(\mathbf{k},t)&=&\frac{1}{N}\sum_{j=1}^N R(\rho_j)e^{i(\mathbf{k}_0-\mathbf{k})\cdot\mathbf{r}_j(t)+i\ell\phi_j(t)}\label{bunching}
 \end{eqnarray}
is the 'optical magnetization', or 'bunching factor'. The direction of the scattered field is determined by the wave-vector $\mathbf{k}$ in Eq.(\ref{bunching}), which depends on the spatial distribution of the atoms.

\section{Atoms in a transverse plane}

We are interested in the case where the atoms are confined in the transverse plane, with $z_j=0$. Then Eq.(\ref{CARL:pj}) becomes
\begin{eqnarray}
\dot \mathbf{p}_j &=&
-\Gamma\hbar k_0\left(\frac{\Omega_0}{2\Delta_0}\right)^2R(\rho_j)\sum_{m\neq j}R(\rho_m)\left\{
\frac{\sin[k_0\rho_{jm}-\ell\phi_{jm}]}{k_0\rho_{jm}}+\frac{\cos[k_0\rho_{jm}-\ell\phi_{jm}]}{(k_0\rho_{jm})^2}
\right\}\mathbf{\hat u}_{jm}\nonumber\\
&+&
\hbar\Gamma\left(\frac{\Omega_0}{2\Delta_0}\right)^2R(\rho_j)\left[\frac{|\ell|}{\rho_j}-\frac{2\rho_j}{w^2}\right]
\sum_{m\neq j}R(\rho_m)\frac{\cos[k_0\rho_{jm}-\ell\phi_{jm}]}{k_0\rho_{jm}}\mathbf{\hat u}_\rho\nonumber\\
&+&\ell\hbar\Gamma\left(\frac{\Omega_0}{2\Delta_0}\right)^2\frac{R(\rho_j)}{\rho_j}\sum_{m=1}^NR(\rho_m)\frac{\sin[k_0\rho_{jm}-\ell\phi_{jm}]}{k_0\rho_{jm}}\mathbf{\hat u}_\phi+\mathbf{F}_{\rho j},
\label{CARL:2D}
\end{eqnarray}
where $\mathbf{\hat u}_{jm}=(\mathbf{r}_{\perp j}-\mathbf{r}_{\perp m})/\rho_{jm}$, with $\rho_{jm}=\sqrt{\rho_j^2+\rho_m^2-2\rho_j\rho_m\cos\phi_{jm}}$ and $\mathbf{r}_{\perp j}=\rho_j\cos\phi_j\mathbf{\hat u}_{x}+\rho_j\sin\phi_j\mathbf{\hat u}_{y}$.
The last term in Eq.(\ref{CARL:2D}) is the single-atom radial force exerted by the pump,
\begin{equation}
\mathbf{F}_{\rho j}=\frac{\hbar[\Omega_0R(\rho_j)]^2}{2\Delta_0}\left(\frac{\ell}{\rho_j}-\frac{2\rho_j}{w^2}\right)\mathbf{\hat u}_\rho. 
\end{equation}
We have neglected the longitudinal radiation pressure force $\mathbf{F}_{z j}$ and included the azimuthal force $\mathbf{F}_{\phi j}$ in the last sum over $m$ of Eq.(\ref{CARL:2D}), with the term $m=j$.
This is a rather surprising result which will have important consequences in the occurrence of the azimuthal CARL instability, and it will be discussed in section V.

By projecting Eq.(\ref{CARL:2D}) along the radial and azimuthal directions, we obtain
\begin{eqnarray}
\ddot \rho_j&=& \sum_{m\neq j}\frac{f_{jm}}{\rho_{jm}}(\rho_j-\rho_m\cos\phi_{jm})+h_j+\rho_j\dot\phi_j^2\label{force:rho}, \\
2\dot\rho_j\dot\phi_j+\rho_j\ddot\phi_j &=& \sum_{m\neq j}\frac{f_{jm}}{\rho_{jm}}\rho_m\sin\phi_{jm}+g_j, \label{force:phi}
\end{eqnarray}
with
\begin{eqnarray}
f_{jm}&=&-\Gamma\frac{\hbar k_0}{M}\left(\frac{\Omega_0}{2\Delta_0}\right)^2 R(\rho_j)R(\rho_m)\left\{
\frac{\sin(k_0\rho_{jm}-\ell\phi_{jm})}{k_0\rho_{jm}}+\frac{\cos(k_0\rho_{jm}-\ell\phi_{jm})}{(k_0\rho_{jm})^2}\right\}\label{fjm},\\
h_{j}&=& \frac{\hbar\Gamma}{M}\left(\frac{\Omega_0}{2\Delta_0}\right)^2 R(\rho_j)\left[\frac{\ell}{\rho_j}-\frac{2\rho_j}{w^2}\right]\left\{
\sum_{m\neq j}R(\rho_m)\frac{\cos(k_0\rho_{jm}-\ell\phi_{jm})}{k_0\rho_{jm}}
+\frac{2\Delta_0}{\Gamma}R(\rho_j)\right\}\label{hj},\\
g_{j}&=&\ell \frac{\hbar\Gamma}{M}\left(\frac{\Omega_0}{2\Delta_0}\right)^2\frac{R(\rho_j)}{\rho_j}\sum_{m=1}^NR(\rho_m)\frac{\sin(k_0\rho_{jm}-\ell\phi_{jm})}{k_0\rho_{jm}}.\label{gj}
\end{eqnarray}
We observe that the first term on the r.h.s. of Eq.(\ref{CARL:2D}) gives both a radial and azimuthal contribution to the force (term $f_{jm}$ in Eqs.(\ref{force:rho}) and (\ref{force:phi})), whereas its second term and the pump gradient force $\mathbf{F}_{\rho j}$ give a purely radial force (term $h_j$ in Eq.(\ref{force:rho})). Finally the third term of the r.h.s. of Eq.(\ref{CARL:2D}) gives an azimuthal force proportional to the  index $\ell$ (term $g_j$ in Eq.(\ref{force:phi})).

Considering the emission only in the transverse plane, with $\mathbf{k}=k_0\cos\theta\,\mathbf{\hat u}_{x}+k_0\sin\theta\,\mathbf{\hat u}_{y}$,
the optical magnetization becomes
\begin{eqnarray}
M(\theta)&=&\frac{1}{N}\sum_{j=1}^N R(\rho_j)e^{-ik_0\rho_j\cos(\theta-\phi_j)+i\ell\phi_j}\label{bunching:phi}.
 \end{eqnarray}
We observe that its phase is modulated by the different atomic positions in the transverse plane.
Using the identity
\begin{equation}
e^{iz\cos\phi}=\sum_{m=-\infty}^{+\infty}i^m J_m(z)e^{im\phi},\nonumber
\end{equation}
where $J_m(z)$ is the $m$-order Bessel function, we can write
\begin{eqnarray}
M(\theta)&=&\sum_{m=-\infty}^{+\infty}P_{\ell,m}e^{im\theta},
 \end{eqnarray}
where
\begin{equation}
 P_{\ell,m}=(-i)^m\frac{1}{N}\sum_{j=1}^N R(\rho_j)J_m(k_0\rho_j)e^{i(\ell-m)\phi_j}.
\end{equation}
Hence, the angular distribution of the scattered intensity is determined by the radial and azimuthal atomic positions.

\section{Atoms on a ring}

Let us consider the case where the atoms are trapped in a thin circular
ring in the transverse plane such that the radial motion of the
atoms is negligible, i.e. $\rho_j=\rho $ is constant. Then, the evolution of the atomic motion is
essentially one-dimensional in the azimuth angle  $\phi_j$. By considering $\rho_{jm}=2\rho|\sin(\phi_{jm}/2)|$, together with Eqs.(\ref{force:rho}) and (\ref{force:phi}), we obtain the azimuthal equation 
\begin{eqnarray}
\ddot\phi_j &=& \omega_r\Gamma\left(\frac{\Omega_0}{2\Delta_0}\right)^2
\frac{R^2(\rho)}{(k_0\rho)^2}\left\{
-\sum_{m\neq j}\cot(\phi_{jm}/2)\left[
\sin[2k_0\rho|\sin(\phi_{jm}/2)|-\ell\phi_{jm}]+\frac{\cos[2k_0\rho|\sin(\phi_{jm}/2)|-\ell\phi_{jm}]}{2k_0\rho|\sin(\phi_{jm}/2)|}
\right]\right.\nonumber\\
&+&\left.\ell\sum_{m=1}^N\frac{\sin[2k_0\rho|\sin(\phi_{jm}/2)|-\ell\phi_{jm}]}{k_0\rho|\sin(\phi_{jm}/2)|}\right\}\label{phases}
\end{eqnarray}
We observe that, if $\rho= w/\sqrt{2}=1/k_\theta$, the torque on the $j$-atom is proportional to the angular recoil frequency $\omega_\theta=\hbar k_\theta^2/2M$ rather than to the recoil frequency $\omega_r$.
When $\rho_j$ is constant the optical magnetization can be written as
\begin{equation}
M(\theta)=\frac{R(\rho)}{N}\sum_{j=1}^N e^{-ik_0\rho\cos(\theta-\phi_j)+i\ell\phi_j}=
R(\rho)\sum_m (-i)^m J_m(k_0\rho)\Phi_{m-\ell}e^{im\theta},\label{M}
\end{equation}
where
\begin{equation}
\Phi_{n}=\frac{1}{N}\sum_{j=1}^N e^{-in\phi_j}
\end{equation}
is the azimuthal bunching on the $n$th harmonic. In the case of a uniform distribution of the phases $\phi_j$, its expression becomes 
\begin{eqnarray}
M(\theta)&=&R(\rho)(-i)^\ell J_\ell(k_0\rho)e^{i\ell\theta},
\end{eqnarray}
so that $|M(\theta)|=R(\rho)|J_\ell(k_0\rho)|$ and the scattered intensity is isotropic in the transverse plane. This is true also if the phases are perfectly bunched around a single value, $\phi_j=\phi_0$:
\begin{eqnarray}
M(\theta)&=&R(\rho)e^{i\ell\phi_0-i\rho\cos(\theta-\phi_0)}, 
\end{eqnarray}
with $|M(\theta)|=R(\rho)$. More generally, $M(\theta)$ is a superposition of different harmonics.

\section{Azimuthal OAM-CARL model}

We study the azimuthal motion of the atoms in the 2D transverse plane, described by Eqs.(\ref{phases}). To avoid the singularity for $\phi_{jm}\rightarrow 0$, we introduce a cut-off $\epsilon$ such that
\begin{equation}
|\sin(\phi_{jm}/2)|\rightarrow \sqrt{\sin^2(\phi_{jm}/2)+\epsilon^2}\equiv q_{jm}.
\end{equation}
Furthermore, since $\rho$ is kept constant, we rescale the time $t$ into the dimensionless time $t'=\beta t$, with
\begin{equation}
\beta=\sqrt{\omega_r\Gamma}\left(\frac{\Omega_0}{2\Delta_0}\frac{R(\rho)}{k_0\rho}\right)
\end{equation}
Hence, the working equations are:
\begin{eqnarray}
\frac{d^2\phi_j}{dt'^2} &=&
-\sum_{m\neq j}\frac{\sin(\phi_{jm})}{2q_{jm}^2}\left\{
\sin[2k_0\rho  q_{jm}-\ell\phi_{jm}]+\frac{\cos[2k_0\rho q_{jm}-\ell\phi_{jm}]}{2k_0\rho  q_{jm}}
\right\}+\ell\sum_{m=1}^N\frac{\sin[2k_0\rho  q_{jm}-\ell\phi_{jm}]}{k_0\rho q_{jm}}.
\label{phases:scaled}
\end{eqnarray}
For a single atom, in the limit $\epsilon\rightarrow 0$  the unscaled equation for the single atom  with phase $\phi$ is
\begin{eqnarray}
\ddot\phi &=&2\ell\,\beta^2=\ell\frac{\hbar\Gamma}{M\rho ^2}\left(\frac{\Omega_0 R(\rho)}{2\Delta_0}\right)^2,
\end{eqnarray}
It can be written in terms of the atomic angular momentum along the $z$-axis, $L_z=M\rho^2\dot\phi $:
\begin{equation}
\dot L_z =\ell\hbar\Gamma\left(\frac{\Omega_0 R(\rho)}{2\Delta_0}\right)^2=T_z,
\end{equation}
where $T_z$ is the torque exerted by the OAM pump, in agreement with ref.\cite{babiker1994light}. In the scattering force, the pump photon also transfers, other than its linear momentum $\hbar \mathbf{k}_0$, its angular momentum $(\hbar\ell)\mathbf{\hat z}$, and the torque $T_z$ is equal to the photon angular momentum $\hbar\ell$ times the scattering rate.

Let us write Eqs.(\ref{phases:scaled}) in the form:
\begin{eqnarray}
\frac{d^2\phi_j}{dt'^2} &=& \sum_{m=1}^N G(\phi_j-\phi_m)\label{p:j}
\end{eqnarray}
with $j=1,\dots,N$ and where
\begin{equation}
G(x)=-\frac{\sin x}{2q(x)^2}\left[
\sin[2\rho'  q(x)-\ell x]+\frac{\cos[2\rho' q(x)-\ell x]}{2\rho'  q(x)}
\right]+\ell\frac{\sin[2\rho' q(x)-\ell x]}{\rho' q(x)},\label{force}
\end{equation}
$\rho'=k_0\rho$, and
\begin{equation}
q(x)=\sqrt{\sin^2(x/2)+\epsilon^2}.
\end{equation}
An important feature is that the force $G(\phi_j-\phi_m)$ on particle $j$ due to particle $m$, does not have the symmetry of a force derivable from a two-body interaction potential, which is a function solely of the separation between particles, i.e., $G(\phi_j-\phi_m)\neq -G(\phi_m-\phi_j)$. As a consequence, the average angular velocity $\langle\dot\phi\rangle=(1/N)\sum_{j=1}^N\dot\phi_j$ is not conserved in time. The dynamics of Eqs.(\ref{p:j}) are not derivable from an underlying Hamiltonian, so that one may not associate an energy function with the system \cite{bachelard2019slow}.

The function $G(x)$ can be derived from a potential $V(x)$,
\begin{equation}
G(x)=-\frac{d}{dx}V(x),
\end{equation}
where
\begin{equation}
V(x)=-\frac{\cos[2\rho' q(x)-\ell x]}{\rho' q(x)}.\label{potential}
\end{equation}
Therefore, Eq.(\ref{p:j}) can be rewritten as:
\begin{eqnarray}
\frac{d^2\phi_j}{dt'^2} &=& -\frac{\partial}{\partial\phi_j}\sum_{m=1}^N V(\phi_j-\phi_m).\label{p:V}
\end{eqnarray}

\subsection{Equilibrium}
In the case the phases are uniformly distributed (no bunching), we consider $\phi_j$ as continuous variables $\phi$ and we approximate the sum in Eq.(\ref{p:V}) by an integral:
\begin{eqnarray}
\frac{d^2\phi}{dt'^2} &=& -\frac{N}{2\pi}\frac{\partial}{\partial\phi}\int_0^{2\pi}d\phi' V(\phi-\phi')=
 \frac{N}{2\pi}\int_0^{2\pi}d\phi'\frac{\partial V(\phi-\phi')}{\partial\phi'} 
\end{eqnarray}
Changing the integration variable into $x=\phi-\phi'$ and using the fact that $V(x)$ is periodic between $0$ and $2\pi$, we can write
\begin{eqnarray}
\frac{d^2\phi}{dt'^2}  &=& 
 -\frac{N}{2\pi}\int_{\phi}^{\phi+2\pi}dx\frac{dV(x)}{dx} = -\frac{N}{2\pi}[V(\phi+2\pi)-V(\phi)]=0
\end{eqnarray}
Hence, if the distribution of the phases is uniform, the force on every atoms is zero and the system is in equilibrium. The system is unstable under small perturbations of this initial conditions, with a rate proportional to $\sqrt{N}$ and a lethargy time proportional to $\ln(N)$, as it will be proved in the next section.

\subsection{Stability Analysis}

\begin{figure}[hbt!]
 \centerline{\includegraphics[width=0.5\textwidth]{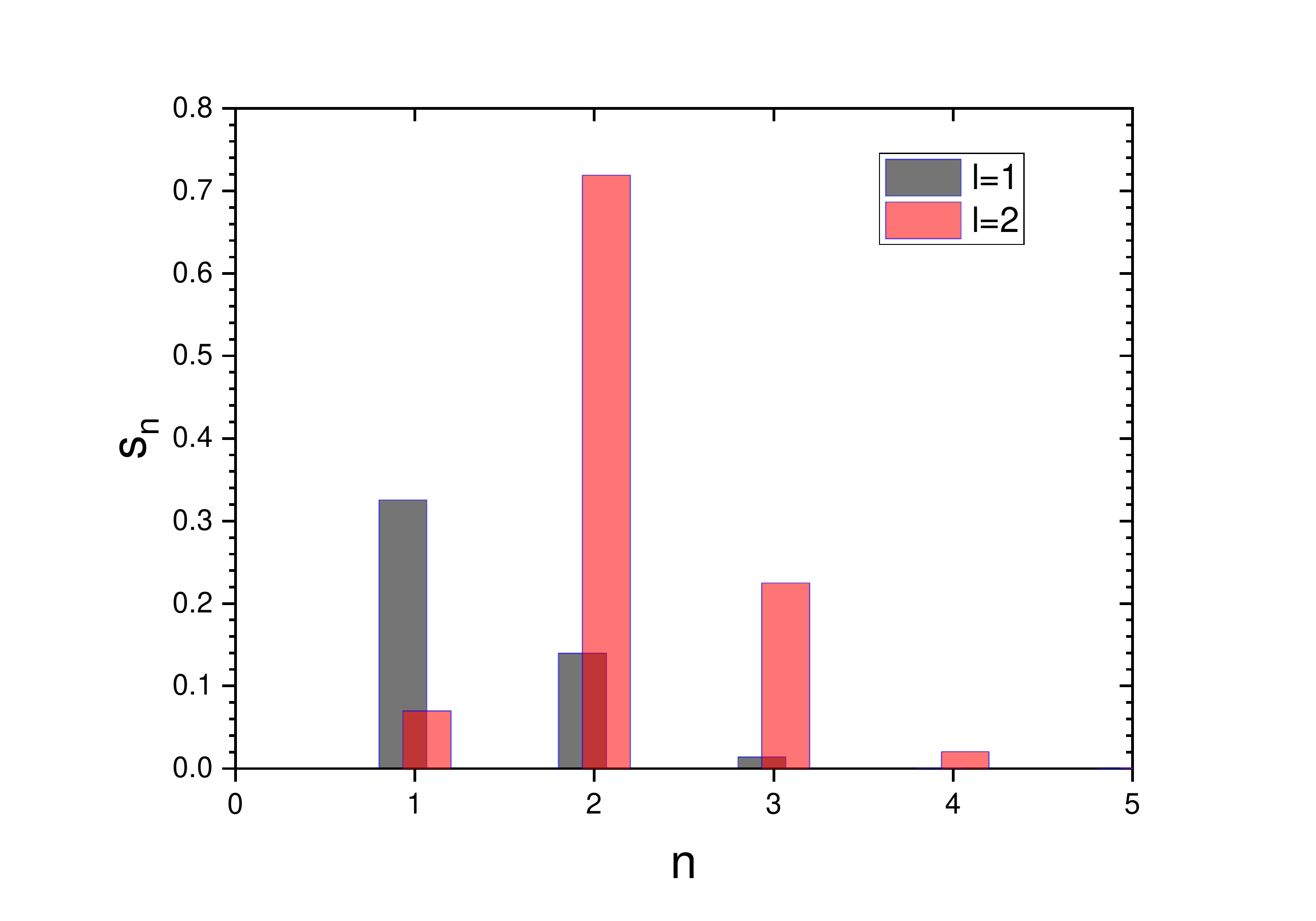}}
\caption{$s_n$ vs. $n$ for $\ell=1$ and $\ell=2$, $k_0\rho=1$ and $\epsilon=0.1$.}
\label{fig:2} 
\end{figure}

\begin{figure}[hbt!]
 \centerline{\includegraphics[width=0.5\textwidth]{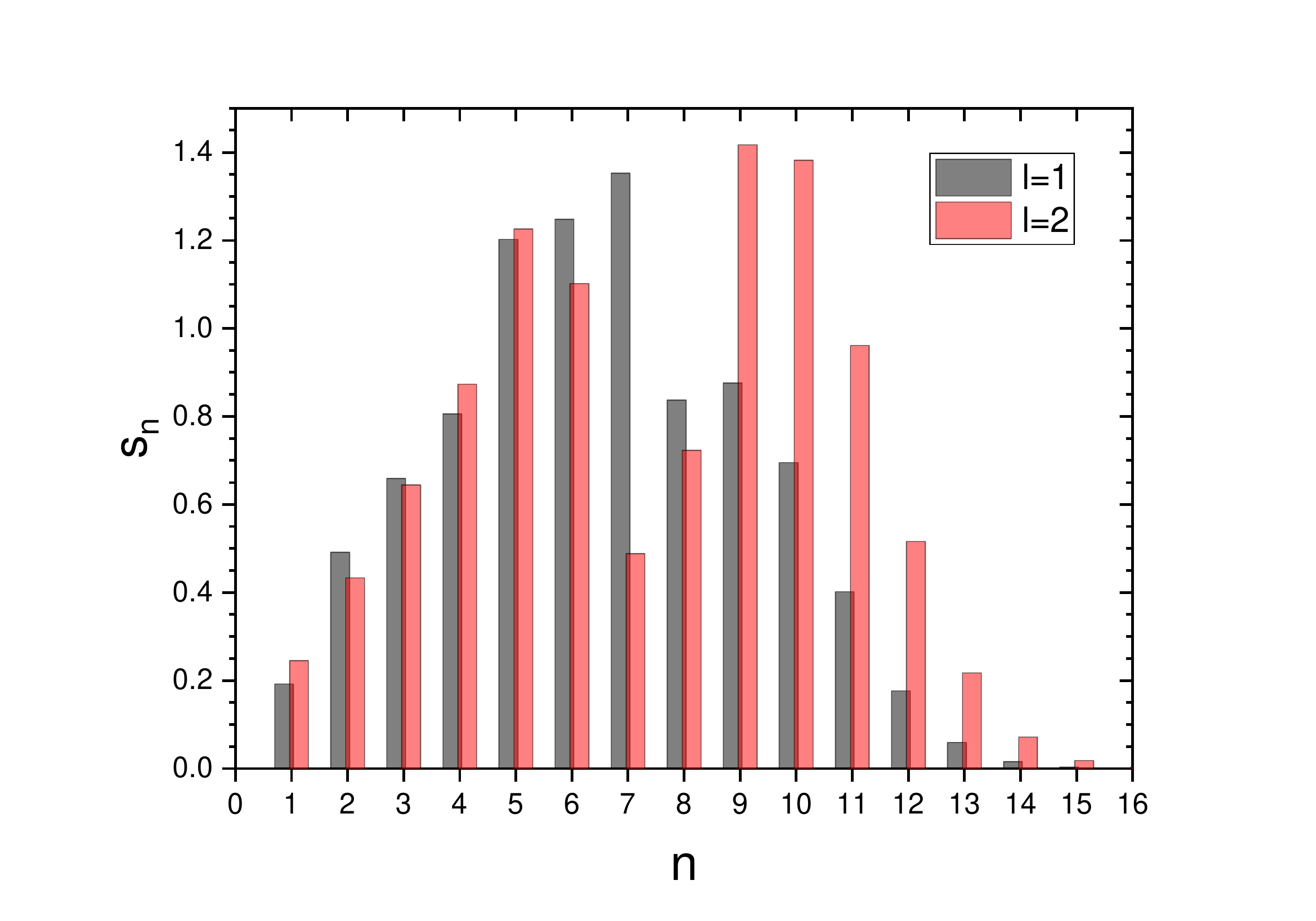}}
\caption{$s_n$ vs. $n$ for $\ell=1$ and $\ell=2$, $k_0\rho=10$ and $\epsilon=0.1$.} 
\label{fig:3}
\end{figure}

We have proved that the azimuthal motion equations Eqs.(\ref{p:j}) have an equilibrium for $\phi_j=\phi_j^{(0)}$, such that
\begin{equation}
\sum_m G\left(\phi_j^{(0)}-\phi_m^{(0)}\right)=0
\end{equation}
and $\dot\phi_j=0$. Let us perturb the equilibrium, $\phi_j(t)=\phi_j^{(0)}+\delta\phi_j(t)$, with $\delta\phi_j\ll\phi_j^{(0)}$. Then, 
the linearized $n$th-harmonic azimuthal bunching
\begin{equation}
\delta\Phi_n(t')=\frac{1}{N}\sum_j e^{-in\phi_j^{(0)}}\delta\phi_j(t'),
\end{equation}
grows in time as $\Phi_n(t')\propto e^{\lambda_n t'}$ (see Appendix \ref{App:3}), where
\begin{equation}
\lambda_n=\pm i n\sqrt{N V_n},\label{eigenvalue}
\end{equation}
and where $V_n$ is the Fourier transform of the potential $V(x)$:
\begin{eqnarray}
V_n&=&\frac{1}{2\pi}\int_0^{2\pi}V(x)e^{-inx}dx.
\end{eqnarray}
Hence, the azimuthal superradiant rate for the $n$th harmonic bunching is
 \begin{equation}
  \Gamma_{n}=\sqrt{N}\beta s_n(k_0\rho,\ell)=\sqrt{\omega_r\Gamma N}\left(\frac{\Omega_0}{2\Delta_0}\frac{R(\rho)}{k_0\rho}\right)
  s_n,
  \end{equation}
where
\begin{equation}
s_n=n\left|\mathrm{Im}\sqrt{V_n}\right|.
\end{equation}
The corresponding frequency shift is
 \begin{equation}
  \delta\Omega_{n}=\sqrt{N}\beta \delta\omega_n=
  \sqrt{\omega_r\Gamma N}\left(\frac{\Omega_0}{2\Delta_0}\frac{R(\rho)}{k_0\rho}\right)
  \delta\omega_n,
  \end{equation}
with
\begin{equation}
\delta\omega_n=n\,\mathrm{Re}\sqrt{V_n}.
\end{equation}
We observe that the superradiant rate scales as $\sqrt{N}$, typical for superradiance in a classical system \cite{robb2005semiclassical}.
Fig.\ref{fig:2}  shows $s_n$ as a function of $n$ for $k_0\rho=1$, $\ell=1$ and $\ell=2$. The cut-off is set to $\epsilon=0.1$. We observe that for a very focused beam ($k_0\rho\le 1$) the maximum bunching occurs for the harmonic $n=\ell$.  However, things become more complex when the ring radius is larger, for  $k_0\rho\gg 1$. Fig.\ref{fig:3} shows $s_n$ as a function of $n$ for $k_0\rho=10$, with $\ell=1$ and $\ell=2$: in these cases the harmonic distribution is wider, with a single maximum around $n=7$ for $\ell=1$ and two relative maxima around $n=5$ and $n=10$ for $\ell=2$.

\section{Numerical results}

Eqs.(\ref{phases:scaled}) have been solved numerically for different values of $\rho$ and $\ell$. The parameters assumed have been $N=100$ and $\epsilon=1$, in order to avoid numerical instabilities occurring when two particles become too close each other. In a real system, with many atoms distributed in larger volumes, this issue should be less critical, with only few particles kicked off by pair collisions. The results presented here, in a idealized situation with the atoms with negligible thermal motion and distributed on a ring of zero thickness, has the objective to show the typical features of the collective effects observable by a laser beam possessing OAM beam incident on the atoms confined in a transverse annular volume. An extension of the analysis including also the radial dynamics and the temperature effects will presented elsewhere.

\begin{figure}[hbt!]
\centerline{\includegraphics[width=0.7\textwidth]{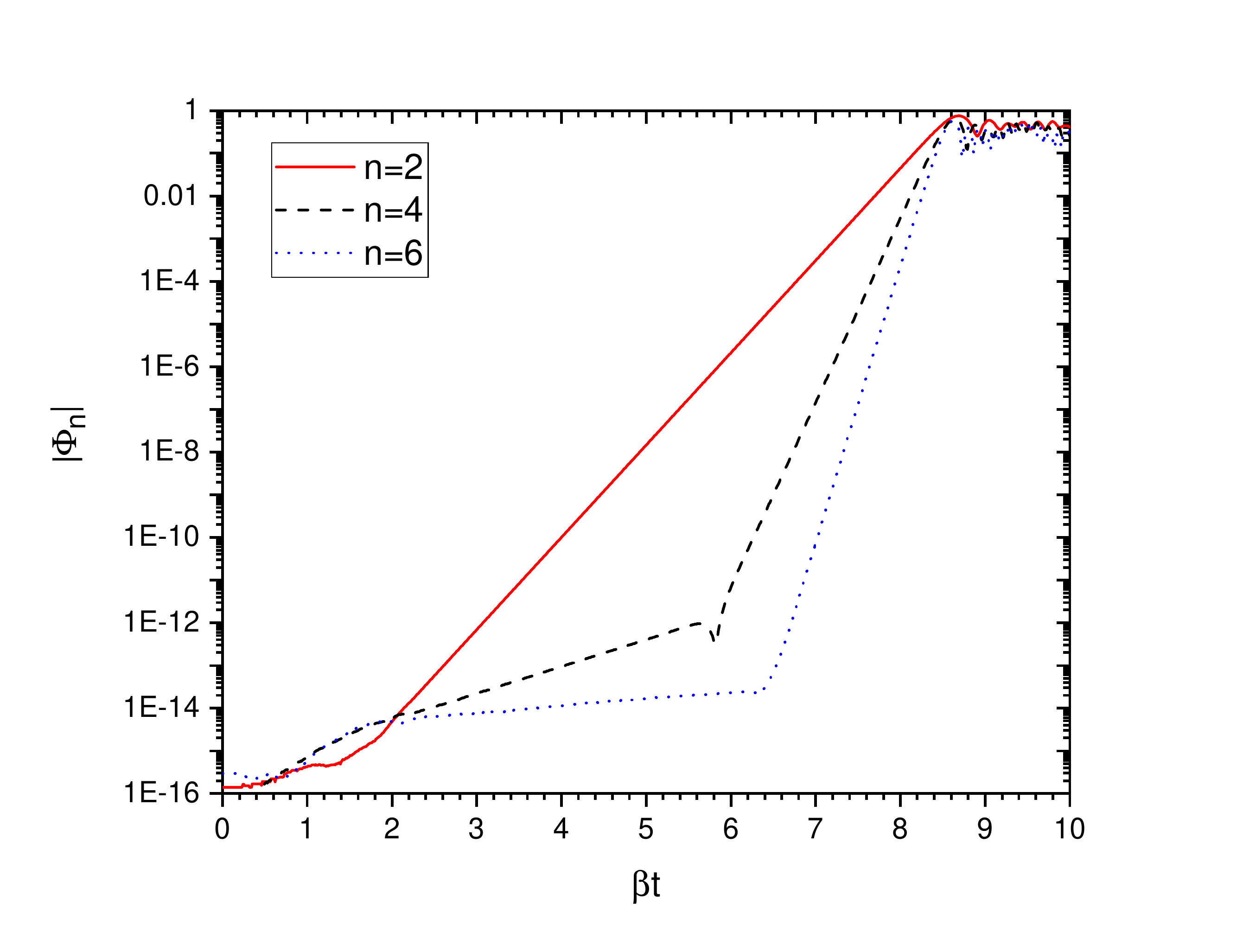}}
\caption{Bunching $|\Phi_n|$ vs. time $\beta t$, for $k_0\rho=1$ and $\ell=1$. The other parameters of the simulation are $N=100$ and $\epsilon=1$.}\label{fig:4}
\end{figure}

\begin{figure}[hbt!]
\centerline{\includegraphics[width=0.7\textwidth]{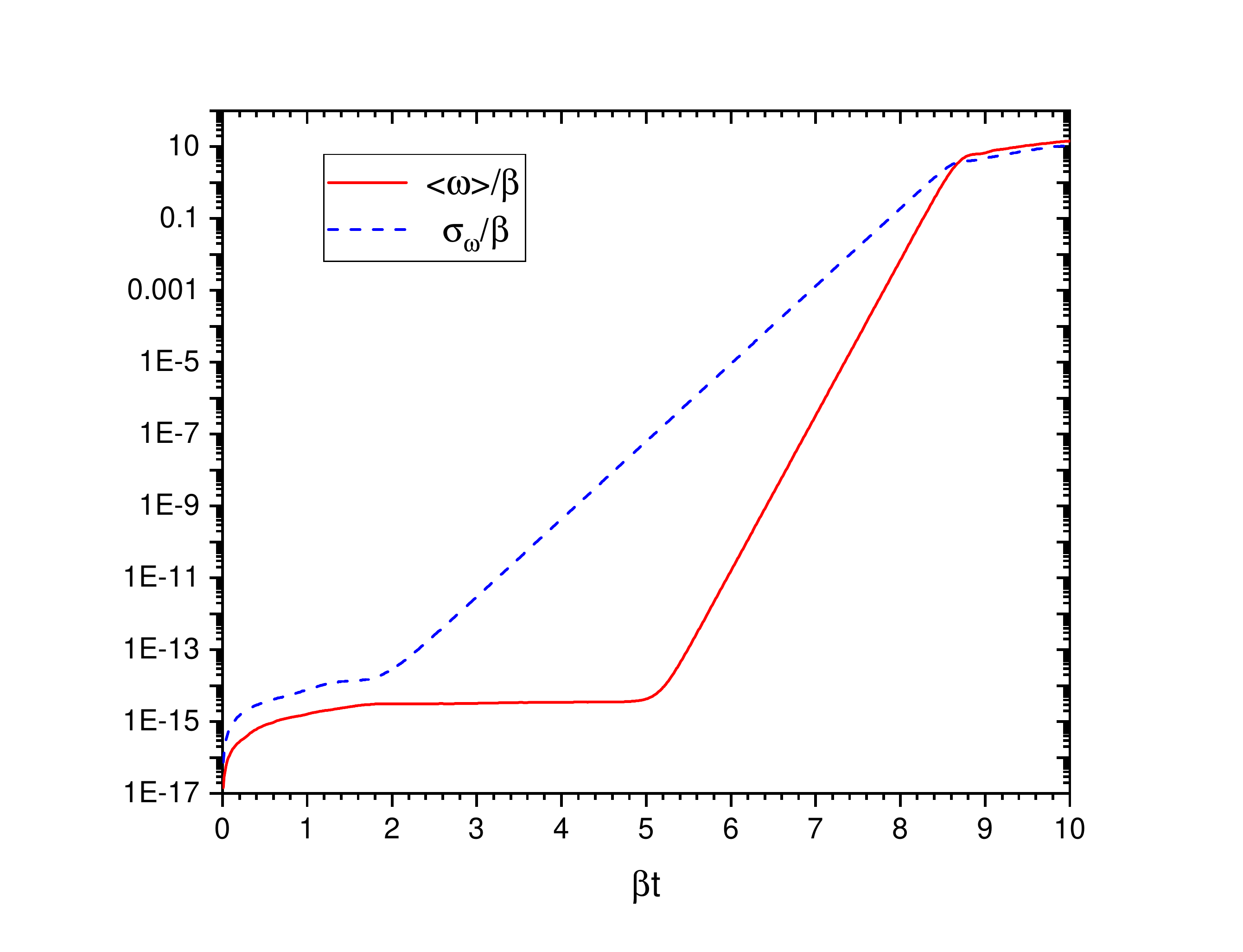}}
\caption{Average angular velocity $\langle\omega\rangle/\beta$ and spread $\sigma_\omega/\beta=\sqrt{\langle\omega^2\rangle-\langle\omega\rangle^2}/\beta$ vs. time $\beta t$, for $k_0\rho=1$ and $\ell=1$.}\label{fig:5}
\end{figure}

\begin{figure}[hbt!]
\centerline{\includegraphics[width=1.0\textwidth]{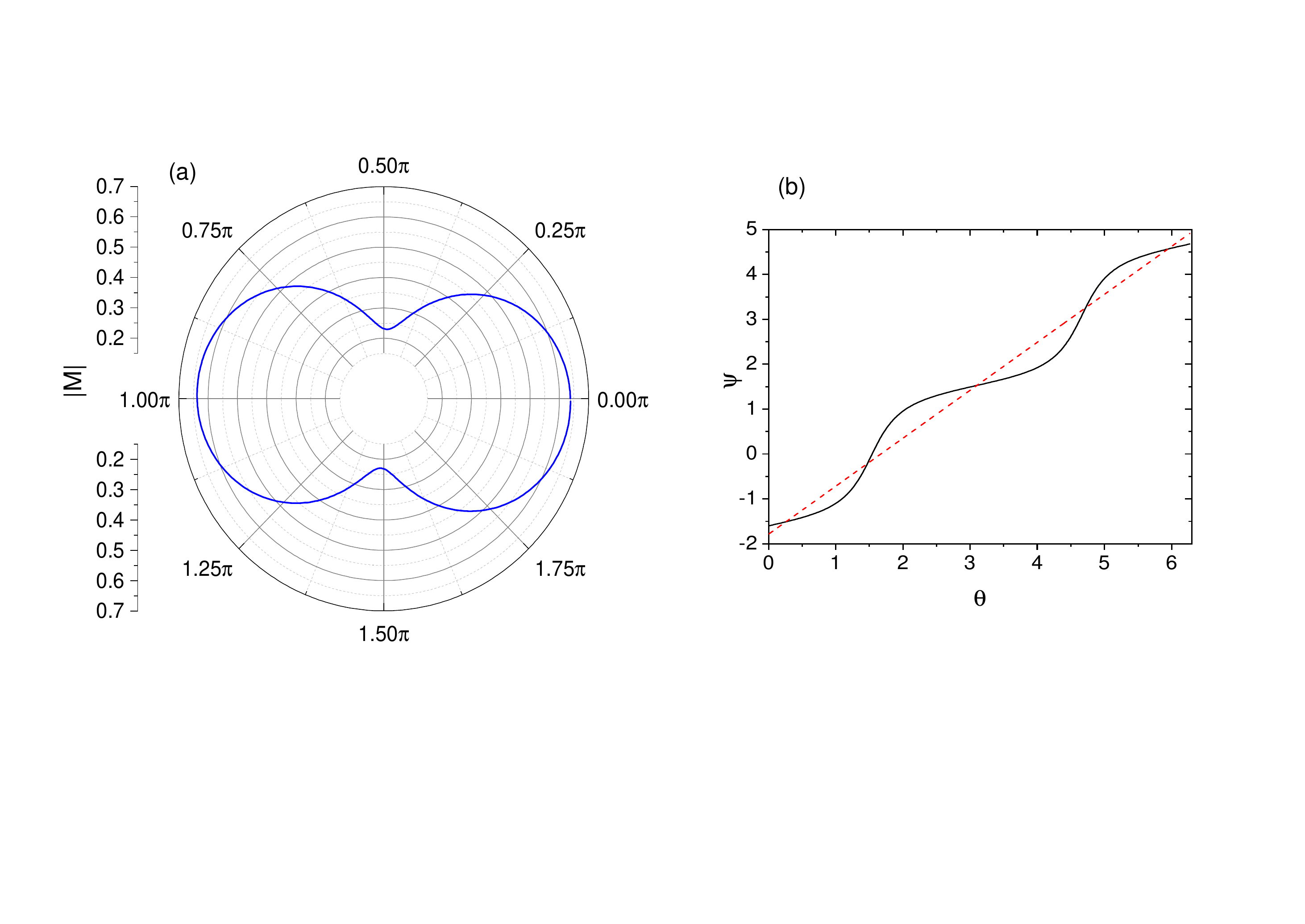}}
\caption{Optical magnetization $M(\theta)=|M(\theta)|\exp(i\psi)$ for $k_0\rho=1$ and $\ell=1$, at $\beta t=8.5$. (a) Polar plot of $|M(\theta)|$; (b) Phase $\psi$ vs. $\theta$. The dashed line is a linear fit with $\psi=-1.78+1.06\,\theta$.}\label{fig:6}
 \end{figure}

\begin{figure}[hbt!]
\centerline{\includegraphics[width=0.7\textwidth]{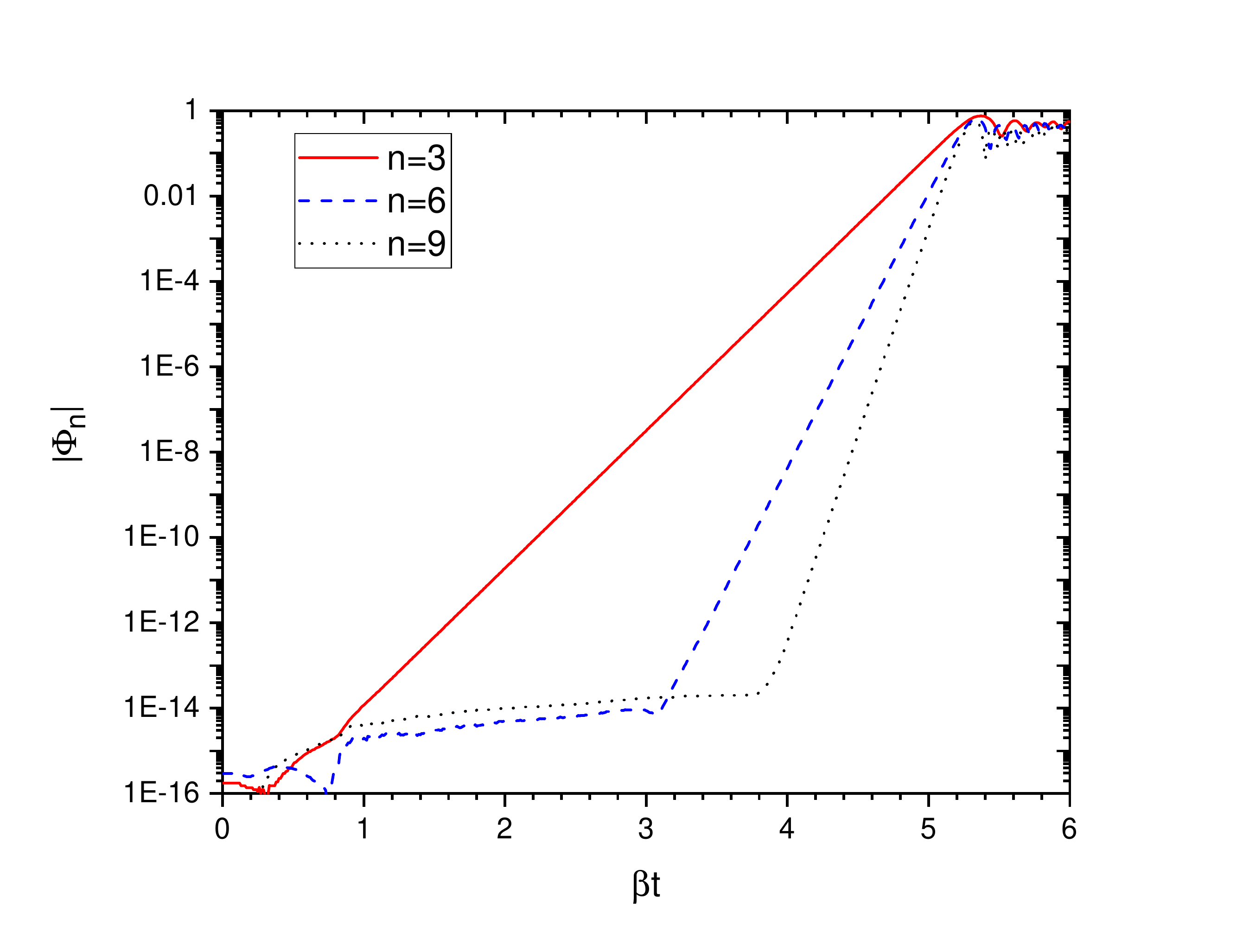}}
\caption{Bunching $|\Phi_n|$ vs. time $\beta t$, for $k_0\rho=1$ and $\ell=2$.}\label{fig:7}
 \end{figure}

\begin{figure}[hbt!]
\centerline{\includegraphics[width=1.0\textwidth]{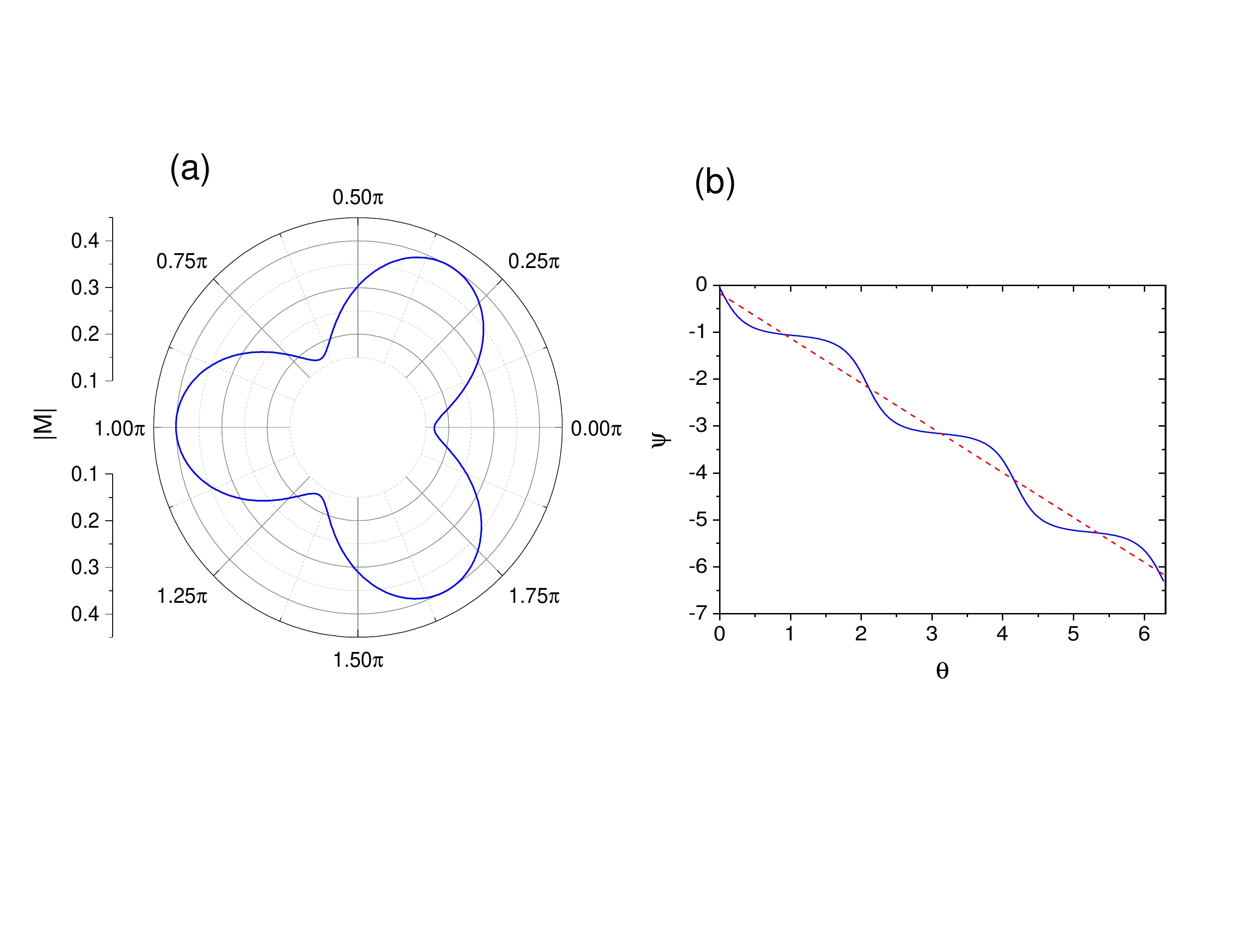}}
\caption{Optical magnetization $M(\theta)=|M(\theta)|\exp(i\psi)$ for $k_0\rho=1$ and $\ell=2$, at $\beta t=8.5$. (a) Polar plot of $|M(\theta)|$; (b) Phase $\psi$ vs. $\theta$. The dashed line is a linear fit with $\psi=-0.17-0.95\,\theta$.}\label{fig:8}
\end{figure}

First, we have considered a case of a tightly focused laser beam, with $k_0\rho=1$. Fig.\ref{fig:4} shows the azimuthal bunching $|\Phi_n|$ vs. time for $\ell=1$. The growth occurs for $n=2$ and its first harmonics $n=4,6$, up to a value close to unity, corresponding to the atoms bunched around two opposite phases, with difference $\pi$. The atoms start to rotate counterclockwise when the bunching  $|\Phi_2|$ becomes significant, as can be seen in Fig.\ref{fig:5}, showing the average angular velocity $\langle\omega\rangle/\beta$ vs. time (where $\omega_j=d\phi_j/dt$) (full red line) and the spread of the angular velocity $\sigma_\omega/\beta=\sqrt{\langle\omega^2\rangle-\langle\omega\rangle^2}/\beta$. The phases are accelerated up to an angular velocity of about $10\beta$. Fig.\ref{fig:6} shows the polar plot of $|M(\theta)|$, (a), and its phase $\psi$, (b), vs $\theta$ (where $M(\theta)=|M(\theta)|\exp[i\psi(\theta)]$). The emission occurs in the transverse plane within two lobes, in a typical bipolar distribution, rotating with the atomic average angular velocity. We observe that the scattered field is, from Eq.(\ref{M}),
\begin{equation}
M(\theta)\sim
R(\rho) e^{i\theta}\left\{J_1(k_0\rho)\left[1-i\Phi_{2}^*e^{-2i\theta}\right]-J_3(k_0\rho)\Phi_{2}e^{2i\theta}\right\}
\end{equation}
Since $J_3(k_0\rho)\ll J_1(k_0\rho)$ and $|\Phi_2|<1$, the phase is $\psi(\theta)\sim\theta_0+\theta$ with a modulation $\propto\sin(2\theta+\theta_{02})$ (see Fig.\ref{fig:6}b).
A similar behavior is observed for $k_0\rho=1$ and $\ell=2$, in Figures \ref{fig:7} and \ref{fig:8}: now the angular bunching is for $n=3$ and its harmonics and the field angular distribution has three symmetric lobes (see Fig.\ref{fig:8}a), meaning that the atoms bunch around three phases separated by $2\pi/3$. The field in this case is
\begin{equation}
M(\theta)\sim
iR(\rho) e^{-i\theta}\left[J_1(k_0\rho)\Phi_{3}^*+iJ_2(k_0\rho)e^{3i\theta}-J_5(k_0\rho)\Phi_{3}e^{6i\theta}\right]
\end{equation}
Since $J_5(k_0\rho)\ll J_2(k_0\rho)\ll J_1(k_0\rho)$ and $|\Phi_3|<1$, the phase is now $\psi(\theta)\sim\theta_0-\theta$ with a modulation as $\sin(3\theta+\theta_{03})$ (see Fig.\ref{fig:8}b).

\begin{figure}[hbt!]
\centerline{\includegraphics[width=0.7\textwidth]{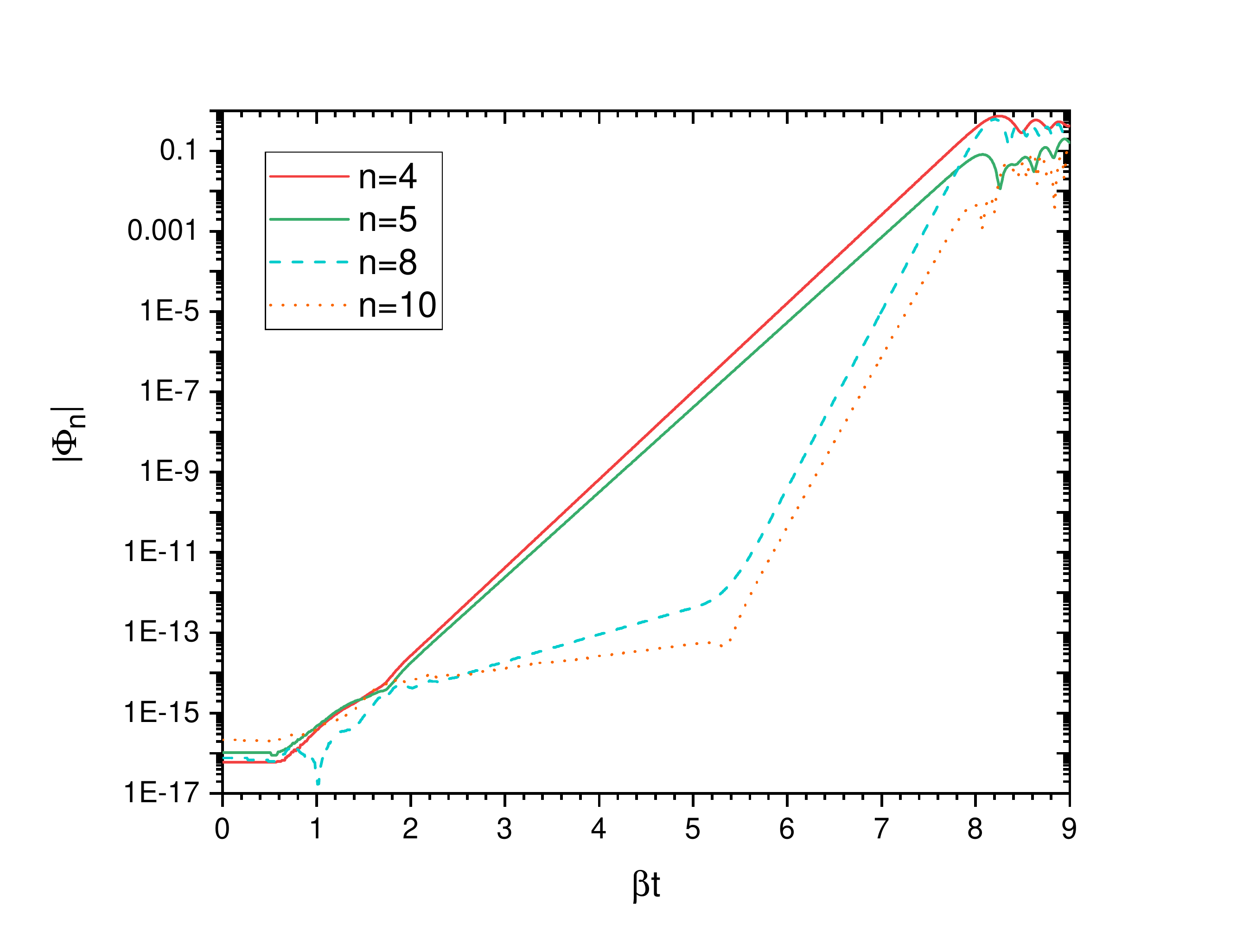}}
\caption{Bunching $|\Phi_n|$ vs. time $\beta t$, for $k_0\rho=10$ and $\ell=1$.}\label{fig:9}
 \end{figure}

\begin{figure}[hbt!]
\centerline{\includegraphics[width=0.7\textwidth]{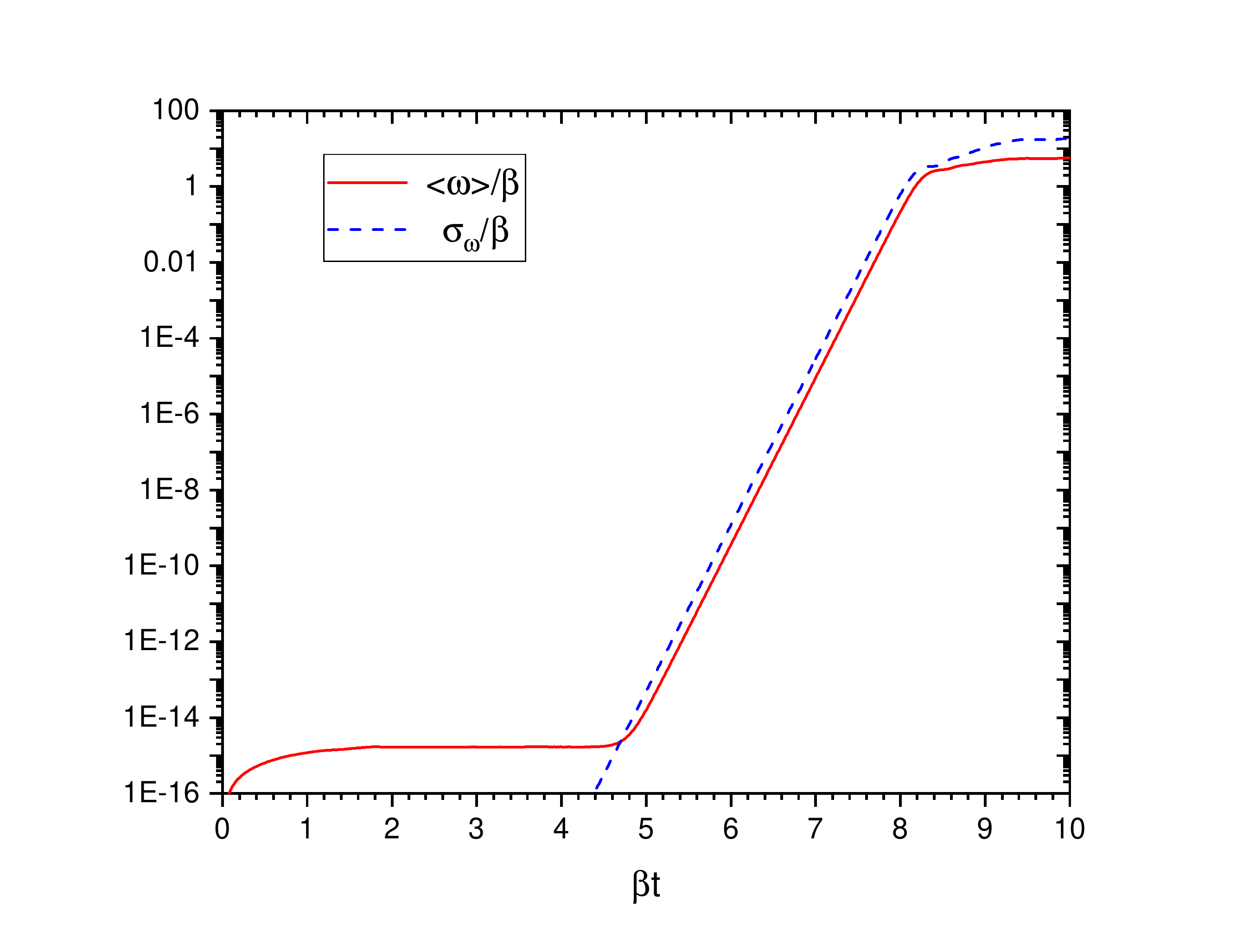}}
\caption{Average angular velocity $\langle\omega\rangle/\beta$ and spread $\sigma_\omega/\beta=\sqrt{\langle\omega^2\rangle-\langle\omega\rangle^2}/\beta$ vs. time $\beta t$, for $k_0\rho=10$ and $\ell=1$.}\label{fig:10}
\end{figure}

\begin{figure}[hbt!]
\centerline{\includegraphics[width=1.0\textwidth]{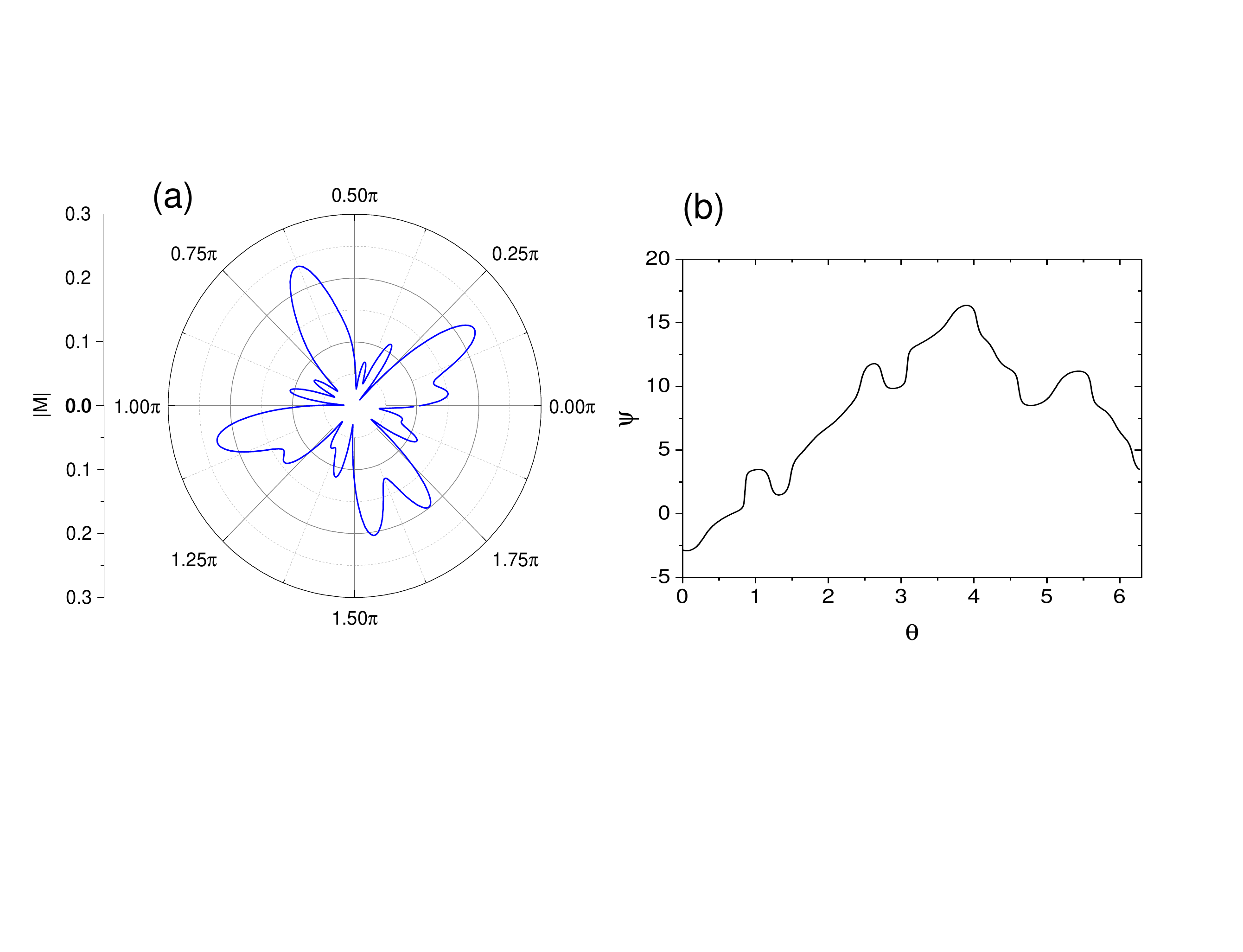}}
\caption{Optical magnetization $M(\theta)=|M(\theta)|\exp(i\psi)$ for $k_0\rho=10$ and $\ell=1$, at $\beta t=8$. (a) Polar plot of $|M(\theta)|$; (b) Phase $\psi$ vs. $\theta$.}\label{fig:11}
 \end{figure}

Secondly, we consider a ring with a large radius, $k_0\rho=10$ in the example shown in Fig.\ref{fig:9}-\ref{fig:11}. In this case more than one mode is exponentially amplified (see for instance the result of the linear analysis, Fig.\ref{fig:3}). Fig.\ref{fig:9}, for $\ell=1$, shows that two modes, $n=4$ and $n=5$ with their harmonics, are amplified, with a maximum value larger for the mode $n=4$. Also in this case the atoms are set in rotation with a maximum average angular velocity of $\langle\omega\rangle\sim 10\beta$ (see Fig.\ref{fig:10}). The angular distribution and the phase of the field present a richer structure, as can be seen from Fig.\ref{fig:11}. In particular, we observe the presence of four main lobes (see Fig.\ref{fig:11}a) with a more complex secondary structure. In general, we observe that by increasing the ring radius we excite more and more modes. In the limit of large radius we expect that the emission distribution will tend to be isotropic in the transverse plane.

\section{Conclusions}
We have presented a study of superradiant scattering of an optical field possessing OAM by a cold atomic gas using a multimode model in which no initial assumption about the spatial structure of the scattered field is made. For simplicity we restricted the analysis to the case where the atoms are distributed in a thin ring and the atomic dynamics involves azimuthal motion only. It was shown that a uniform angular distribution of atoms throughout the ring becomes unstable when illuminated with a far-detuned optical pump field, and that the instability is superradiant in character, resulting in spontaneous rotation of the gas and formation of atomic bunches around the ring and scattered light whose phase profile is dependent on the pump OAM index, $\ell$.
A natural extension to the work presented here is a relaxation of some of the assumptions used, e.g., the restriction to solely azimuthal dynamics to include also radial and even longitudinal dynamics to describe more complex spatial structures, and the extension to include polarization effects, as was done for CARL in \cite{gisbert2020multimode}. This would allow study of interactions involving exchange of both OAM and SAM. Additionally, extension from cold, thermal gases to quantum degenerate gases opens up possibilities for new methods for creation of vortices and persistent currents in BECs, in addition to those described in \cite{Tasgin_2011,Das_2016}.

\appendix

\section{Derivation of the multimode equations (\ref{D:r:j})-(\ref{D:A:j})}\label{App:1}

The force acting on the $j$th atom is obtained from the Heisenberg equation:
\begin{equation}
\dot\mathbf{p}_j=\frac{1}{i\hbar}[\mathbf{p}_j,H_L+H_V]=\mathbf{F}_j,
\end{equation}
where $H_L$ and $H_V$ are given by Eqs.(\ref{HL}) and (\ref{HV}). Using cylindrical coordinates, the force $\mathbf{F}_j$ is:
\begin{eqnarray}
\dot\mathbf{F}_{j} &=& \frac{\hbar\Omega_0}{2}\left\{\sigma_j\left[ik_0R(\rho_j)\mathbf{\hat u}_z
-\frac{dR(\rho_j)}{d\rho_j}\mathbf{\hat u}_\rho+i\ell\frac{R(\rho_j)}{\rho_j}\mathbf{\hat u}_\phi\right]e^{-ik_{0}z_j-i\ell\phi_j}-\textrm{h.c.}\right\}\nonumber\\
&+&i\hbar\left\{\sigma_j\sum_{\mathbf{k}}\mathbf{k}\,g_k a_{\mathbf{k}}^\dagger e^{i\delta_kt-i\mathbf{k}\cdot\mathbf{r}_j}-\textrm{h.c.}\right\},\label{Fj}
\end{eqnarray}
with $\delta_k=\omega_k-\omega_0$. The Heisenberg equation for the multimode field operator $a_{\mathbf{k}}$ is:
\begin{eqnarray}
  \dot{a}_{\mathbf{k}} &=& \frac{1}{i\hbar}[a_{\mathbf{k}},H_V]=-ig_ke^{i\delta_k t}\sum_{j=1}^N \sigma_{j}e^{-i \mathbf{k}\cdot
  \mathbf{r}_j}.\label{ak}
\end{eqnarray}
By inserting the expression of $\sigma_j$  Eq.(\ref{s1:adia}) in Eq.(\ref{ak}), assuming $\Delta_0\gg\Gamma$ and neglecting the vacuum field-induced term in Eq.(\ref{s1:adia}), we obtain
\begin{eqnarray}
  \dot a_{\mathbf{k}} &=& -ige^{i\delta_k t}\sum_{j=1}^N
R(\rho_j)e^{i(\mathbf{k}_0-\mathbf{k})\cdot \mathbf{r}_j+i\ell\phi_j} \label{ak:2},
 \end{eqnarray}
with $g=g_k(\Omega_0/2\Delta_0)$. To obtain the force on the $j$-th atom, we insert Eq.(\ref{s1:adia}) in Eq.(\ref{Fj}), assuming $\Delta_0\gg\Gamma$ and neglecting the terms quadratic in the vacuum field $a_{\mathbf{k}}$. A straightforward calculation yields
\begin{eqnarray}
\mathbf{F}_{j}
 &=&i\hbar g R(\rho_j)
 \sum_{\mathbf{k}}(\mathbf{k}_0-\mathbf{k})\left\{a_{\mathbf{k}}
 e^{-i(\mathbf{k}_0-\mathbf{k})\cdot \mathbf{r}_j-i\ell\phi_j-i\delta_k t}-\textrm{h.c.}\right\}\nonumber\\
 &-&
 \hbar g
 \left\{\left[\frac{dR(\rho_j)}{d\rho_j}\mathbf{\hat u}_\rho-i\ell\frac{R(\rho_j)}{\rho_j}\mathbf{\hat u}_\phi\right]\sum_{\mathbf{k}}a_{\mathbf{k}}
 e^{-i(\mathbf{k}_0-\mathbf{k})\cdot \mathbf{r}_j-i\delta_k t}+\textrm{h.c.}\right\}+\mathbf{F}_{Lj},
 \label{force:j}
\end{eqnarray}
where $\mathbf{F}_{Lj}$ is the force exerted by the laser beam, expressed by Eq.(\ref{FL:j}).

\section{Derivation of  the superradiant equations (\ref{CARL:rj})-(\ref{CARL:pj})}\label{App:2}

In free space the light is scattered in the 3D vacuum modes. Following ref.\cite{Ayllon2019}, we eliminate the scattered field by integrating  Eq.(\ref{D:A:j}) to obtain
\begin{equation}
a_{\mathbf{k}}(t) = a_{\mathbf{k}}(0)e^{-i(\omega_k-\omega_0)t}
-igN\int_0^t\rho_{\mathbf{k}_0-\mathbf{k}}(t-\tau)e^{-i(\omega_k-\omega_0)\tau}d\tau,\label{ak:int}
\end{equation}
with
\begin{equation}
\rho_{\mathbf{q}}(t)=\frac{1}{N}\sum_{j=1}^NR(\rho_j)e^{i\ell\phi_j}e^{i\mathbf{q}\cdot\mathbf{r}_j(t)}.\label{rho_q}
\end{equation}
The first term in Eq.(\ref{ak:int}) gives the free electromagnetic field,
i.e., vacuum fluctuations, and the second term is the
radiation field due to Rayleigh scattering.
If Eq.(\ref{ak:int}) is substituted into Eq.(\ref{force:z:j}) for $\mathbf{p}_j$, we obtain:
\begin{eqnarray}
\dot \mathbf{p}_j &=&
\hbar g^2N\sum_{\mathbf{k}}(\mathbf{k}_0-\mathbf{k})\int_0^td\tau R(\rho_j)\left[\rho_{\mathbf{\mathbf{k}_0-\mathbf{k}}}(t-\tau) e^{-i(\mathbf{k}_0-\mathbf{k})\cdot\mathbf{r}_j-i\ell\phi_j}e^{-i(\omega_k-\omega_0)\tau}+\mathrm{h.c.}\right]\nonumber\\
&+&
i\hbar g^2N\sum_{\mathbf{k}}\int_0^td\tau \left\{\left[\frac{dR(\rho_j)}{d\rho_j}\mathbf{\hat u}_\rho-i\ell\frac{R(\rho_j)}{\rho_j}\mathbf{\hat u}_\phi\right]\rho_{\mathbf{\mathbf{k}_0-\mathbf{k}}}(t-\tau) e^{-i(\mathbf{k}_0-\mathbf{k})\cdot\mathbf{r}_j}e^{-i(\omega_k-\omega_0)\tau}+\mathrm{h.c.}\right\}+\mathbf{F}_{Lj},
\label{CARL:sr}
\end{eqnarray}
where the first term of Eq.(\ref{ak:int}) has been neglected. Then, transforming the sum over $\mathbf{k}$ into an integral and using Eq.(\ref{rho_q}), we attain the coming expression:
\begin{eqnarray}
\dot \mathbf{p}_j &=&
\hbar g^2\frac{V_{ph}}{8\pi^3}\sum_{m\neq j}R(\rho_j)R(\rho_m)\left[e^{-i\mathbf{k}_0\cdot(\mathbf{r}_j-\cdot\mathbf{r}_m)-i\ell(\phi_j-\phi_m)}\int_0^td\tau e^{i\omega_0\tau} \int d\mathbf{k}(\mathbf{k}_0-\mathbf{k})e^{i\mathbf{k}\cdot(\mathbf{r}_j-\cdot\mathbf{r}_m)}e^{-ick\tau}+\mathrm{h.c.}\right]\nonumber\\
&+&
i\hbar g^2\frac{V_{ph}}{8\pi^3}\sum_{m\neq j}R(\rho_m)\left\{\left[\frac{dR(\rho_j)}{d\rho_j}\mathbf{\hat u}_\rho-i\ell\frac{R(\rho_j)}{\rho_j}\mathbf{\hat u}_\phi\right]e^{-i\mathbf{k}_0\cdot(\cdot\mathbf{r}_j-\cdot\mathbf{r}_m)-i\ell(\phi_l-\phi_m)}\right.\nonumber\\
&\times & \left.
\int_0^td\tau e^{i\omega_0\tau} \int d\mathbf{k} e^{i\mathbf{k}\cdot(\mathbf{r}_j-\mathbf{r}_m)}e^{-ick\tau}-\mathrm{h.c.}\right\}+\mathbf{F}_{Lj},
\label{CARL:2:sr}
\end{eqnarray}
in which we used the Markov approximation so that $\mathbf{r}_j(t-\tau)\approx \mathbf{r}_j(t)$. We use the results of ref.\cite{Ayllon2019} to write:
\begin{eqnarray}
\int d\mathbf{k}e^{i\mathbf{k}\cdot(\mathbf{r}_j-\mathbf{r}_m)}e^{-ick\tau}&=&\frac{4\pi^2 k_0^2}{c}\frac{1}{ik_0r_{jm}}
\left[\delta(\tau-r_{jm}/c)-\delta(\tau+r_{jm}/c)\right]\\
\int d\mathbf{k}\mathbf{k}e^{i\mathbf{k}\cdot(\mathbf{r}_j-\mathbf{r}_m)}e^{-ick\tau}&=&
\frac{4\pi^2 k_0^3}{c}\left\{
\frac{1}{ik_0r_{jm}}
\left[\delta(\tau-r_{jm}/c)+\delta(\tau+r_{jm}/c)\right]\right.\nonumber\\
&+&\left.\frac{1}{(k_0r_{jm})^2}
\left[\delta(\tau-r_{jm}/c)-\delta(\tau+r_{jm}/c)\right]\right\}\hat\mathbf{r}_{jm};
\end{eqnarray}
being $\mathbf{r}_{jm}=\mathbf{r}_{j}-\mathbf{r}_{m}$, ~$r_{jm}=|\mathbf{r}_{jm}|$ and $\hat\mathbf{r}_{jm}=\mathbf{r}_{jm}/r_{jm}$.
By inserting these expressions into Eq.(\ref{CARL:2:sr}), together with the definitions of $g$ and $\Gamma$, it is straightfoward to arrive to the following expression of the force equation:
\begin{eqnarray}
\dot \mathbf{p}_j &=&
\Gamma\hbar k_0\left(\frac{\Omega_0}{2\Delta_0}\right)^2R(\rho_j)\sum_{m\neq j}R(\rho_m)\left\{(\hat\mathbf{z}-\hat\mathbf{r}_{jm})
\frac{\sin[k_0(r_{jm}-z_{jm})-\ell\phi_{jm}]}{k_0r_{jm}}-\hat\mathbf{r}_{jm}\frac{\cos[k_0(r_{jm}-z_{jm})-\ell\phi_{jm}]}{(k_0r_{jm})^2}
\right\}\nonumber\\
&+&
\hbar\Gamma\left(\frac{\Omega_0}{2\Delta_0}\right)^2\frac{dR(\rho_j)}{d\rho_j}\sum_{m\neq j}R(\rho_m)\frac{\cos[k_0(r_{jm}-z_{jm})-\ell\phi_{jm}]}{k_0r_{jm}}\mathbf{\hat u}_\rho\nonumber\\
&+&\ell\hbar\Gamma\left(\frac{\Omega_0}{2\Delta_0}\right)^2\frac{R(\rho_j)}{\rho_j}\sum_{m\neq j}R(\rho_m)\frac{\sin[k_0(r_{jm}-z_{jm})-\ell\phi_{jm}]}{k_0r_{jm}}\mathbf{\hat u}_\phi+\mathbf{F}_{L j},
\label{CARL:3:sr}
\end{eqnarray}
with $\phi_{jm}=\phi_j-\phi_m$. The first term is due to the collective momentum recoil, weighted by the radial profile of the pump field. The second term is a collective radial force, due to variation of the radial profile of the pump. Notice that 
$dR(\rho_j)/d\rho_j=R(\rho_j)[\ell/\rho_j-2\rho_j/w^2]$. The third term is the collective azimuthal force, due to the winging phase of the OAM pump.

\section{Derivation of eigenvalues of Eq.(\ref{eigenvalue})}\label{App:3}

The equations
\begin{eqnarray}
\ddot\phi_j &=& \sum_{m=1}^N G(\phi_j-\phi_m)
\end{eqnarray}
with $j=1,\dots,N$ have an equilibrium with $\phi_j=\phi_j^{(0)}$ such that
\begin{equation}
\sum_m G\left(\phi_j^{(0)}-\phi_m^{(0)}\right)=0
\end{equation}
and $\dot\phi_j=0$. Let us perturb the equilibrium with $\phi_j(t)=\phi_j^{(0)}+\delta\phi_j(t)$, with $\delta\phi_j\ll\phi_j^{(0)}$, and
\begin{eqnarray}
\ddot{\delta\phi_j} &=& -\sum_{m=1}^N K\left(\phi_j^{(0)}-\phi_m^{(0)}\right)(\delta\phi_j-\delta\phi_m)\label{lin:1},
\end{eqnarray}
where
\begin{equation}
K(x)=-\frac{d}{dx}G(x)=\frac{d^2}{dx^2}V(x).
\end{equation}
The first term on the right side of Eq.(\ref{lin:1}) is zero since
\begin{eqnarray}
\sum_{m=1}^N K\left(\phi_j^{(0)}-\phi_m^{(0)}\right)&\rightarrow &\frac{N}{2\pi}\int_0^{2\pi}d\phi'K(\phi-\phi')=
-\frac{N}{2\pi}\int_{\phi}^{\phi+2\pi}dx \frac{d}{dx}G(x)\nonumber\\
&=&\frac{N}{2\pi}[G(\phi+2\pi)-G(\phi)]=0,
\end{eqnarray}
and because $G(x)$ is periodic in $(0,2\pi)$. The linear stability is governed by the equations
\begin{eqnarray}
\ddot{\delta\phi_j} &=& \sum_{m=1}^N K\left(\phi_j^{(0)}-\phi_m^{(0)}\right)\delta\phi_m. \label{lin:2}
\end{eqnarray}
We write the potential $V(x)$ as
\begin{equation}
V(x)=-\frac{1}{2\rho' q(x)}\left(e^{-2i\rho' q(x)+i\ell x}+c.c.\right)=\alpha(x)e^{i\ell x}+c.c.
\label{VV}
\end{equation}
where
\begin{eqnarray}
\alpha(x)&=&-\frac{e^{-2i\rho' q(x)}}{2\rho' q(x)},
\end{eqnarray}
so that
\begin{eqnarray}
K(x)=V''(x)&=&[\alpha''(x)+2i\ell\alpha'(x)-\ell^2\alpha(x)]e^{i\ell x}+c.c.=\gamma(x)e^{i\ell x}+c.c.\label{gamma}
\end{eqnarray}
Introducing the linearized $n$th-harmonic azimuthal bunching as
\begin{equation}
\delta\Phi_n(t)=\frac{1}{N}\sum_j e^{-in\phi_j^{(0)}}\delta\phi_j(t),
\end{equation}
from Eqs.(\ref{lin:2}) and (\ref{gamma}) it follows that
\begin{eqnarray}
\ddot{\delta\Phi}_n &=&\frac{1}{N}\sum_je^{-i(n-\ell)\phi_j^{(0)}}\sum_m\gamma\left(\phi_j^{(0)}-\phi_m^{(0)}\right)
e^{-i\ell\phi_m^{(0)})}\delta\phi_m\nonumber\\
& &+\frac{1}{N}\sum_j e^{-i(n+\ell)\phi_j^{(0)}}\sum_m\gamma^*\left(\phi_j^{(0)}-\phi_m^{(0)}\right)
e^{i\ell\phi_m^{(0)}}\delta\phi_m .
\end{eqnarray}
We define
\begin{eqnarray}
F_m^n&=&\sum_j\gamma\left(\phi_j^{(0)}-\phi_m^{(0)}\right)e^{-i(n-\ell)\phi_j^{(0)}},\\
G_m^n&=&\sum_j\gamma^*\left(\phi_j^{(0)}-\phi_m^{(0)}\right) e^{-i(n+\ell)\phi_j^{(0)}},
\end{eqnarray}
so that
\begin{equation}
\ddot{\delta\Phi}_n=\frac{1}{N}\sum_mF_m^n
e^{-i\ell\phi_m^{(0)}}\delta\phi_m+\frac{1}{N}\sum_m G_m^n e^{i\ell\phi_m^{(0)}}\delta\phi_m.
\end{equation}
In order to determine $F_m^n$ and $G_m^n$, we transform the sums over $j$ into integrals considering $\phi_j^{(0)}$ and $\phi_m^{(0)}$ as continuous variables with uniform distribution:
\begin{eqnarray}
F^n(\phi)&=&\frac{N}{2\pi}\int_0^{2\pi}\gamma\left(\phi'-\phi\right) e^{-i(n-\ell)\phi'}d\phi'\\
G^n(\phi)&=&\frac{N}{2\pi}\int_0^{2\pi}\gamma^*\left(\phi'-\phi\right) e^{-i(n+\ell)\phi'}d\phi'.
\end{eqnarray}
Changing integration variable into $x=\phi'-\phi$, we obtain
\begin{eqnarray}
F^n(\phi)&=&e^{-i(n-\ell)\phi}\frac{N}{2\pi}\int_{0}^{2\pi}\gamma(x)e^{-i(n-\ell)x}dx=N\gamma_{n-\ell}e^{-i(n-\ell)\phi}\\
G^n(\phi)&=&e^{-i(n+\ell)\phi}\frac{N}{2\pi}\int_{0}^{2\pi}\gamma^*(x)e^{-i(n+\ell)x}dx=N\gamma^*_{-(n+\ell)}e^{-i(n+\ell)\phi},
\end{eqnarray}
where 
\begin{equation}
\gamma_k=\frac{1}{2\pi}\int_{0}^{2\pi}\gamma(x)e^{-ikx}dx
\end{equation}
is independent on $\phi$.
Hence, we obtain in the discrete,
\begin{eqnarray}
F_m^n &=& N\gamma_{n-\ell}e^{-i(n-\ell)\phi_m^{(0)}}\\
G_m^n &=& N\gamma^*_{-(n+\ell)}e^{-i(n+\ell)\phi_m^{(0)}},
\end{eqnarray}
and
\begin{equation}
\ddot{\delta\Phi}_n=N\left[\gamma_{n-\ell}+\gamma^*_{-(n+\ell)}\right]\delta\Phi_n.
\end{equation}
The coefficient $\gamma_k$ is
\begin{equation}
\gamma_k=\frac{1}{2\pi}\int_0^{2\pi}\left[
\alpha''(x)+2i\ell\alpha'(x)-\ell^2\alpha(x)\right]e^{-ikx}dx.
\end{equation}
By integrating by part we obtain
\begin{equation}
\gamma_k=(k+\ell)^2v_k,\label{gamma_k}
\end{equation}
where
\begin{equation}
v_k=-\frac{1}{2\pi}\int_0^{2\pi}\alpha(x)e^{-ikx}dx.\label{vk}
\end{equation}
Assuming $\Phi_n(t)\propto e^{\lambda_n t}$, then 
\begin{equation}
\lambda_n^2=N[\gamma_{n-\ell}+\gamma^*_{-(n+\ell)}].
\end{equation}
Using Eqs.(\ref{gamma_k}), (\ref{vk}) and (\ref{VV}), we obtain
\begin{equation}
\lambda_n=\pm n\sqrt{N}\sqrt{v_{n-\ell}+v_{-(n+\ell)}^*}=\pm i n\sqrt{N V_n}
\end{equation}
where
\begin{eqnarray}
V_n&=&\frac{1}{2\pi}\int_0^{2\pi}V(x)e^{-inx}dx.
\end{eqnarray}
The growth rate and the frequency for the $n$th-bunching are $|\mathrm{Re}(\lambda_n)|$ and $\mathrm{Im}(\lambda_n)$, where the last expression refers to the solution with $\mathrm{Re}(\lambda_n)>0$.

\bibliography{Bibliography}

\end{document}